\documentclass[trackchanges]{aastex62}
 \usepackage{lineno}
 \usepackage{graphicx}
 \graphicspath{ {./Images/} }
 \usepackage{subfigure}
 \usepackage{hyperref}
 \usepackage{natbib}
 \usepackage{times}
 \usepackage{amsmath}
 \usepackage{textcomp}
 \usepackage{longtable}
 \usepackage{mathptmx}
 \usepackage{xcolor}
 \usepackage{booktabs}
 \usepackage[colorinlistoftodos]{todonotes}
 \usepackage{lineno}
 \usepackage{balance}
\submitjournal{ApJ}
\shorttitle{photo-z}
\shortauthors{Sheng et al.}

\def\asec{\ifmmode^{\prime\prime}\else$^{\prime\prime}$\fi}
\def\degs{\ifmmode ^{\circ}\else$^{\circ}$\fi}
\DeclareUnicodeCharacter{2212}{-}

\begin{document}

\title{Discovering High-$z$ BL Lacs Using {\it Swift}/UVOT and SARA Observations with the Dropout Technique}
\email{sheng2@clemson.edu}
\author[0000-0002-3833-1054]{Y. Sheng}
\affil{Department of Physics and Astronomy, Clemson University, SC 29634-0978, U.S.A.\footnote{A member of the SARA Consortium}}

\author[0009-0003-3381-211X]{K. Imam}
\affil{Department of Physics and Astronomy, Clemson University, Kinard Lab of Physics, Clemson, SC 29634-0978, USA}

\author{A. Kaur}
\affil{Department of Physics and Astronomy, Clemson University, SC 29634-0978, U.S.A.\footnote{A member of the SARA Consortium}}

% \author[0000-0002-8979-5254]{M. Rajagopal}
% \affil{Department of Physics and Astronomy, Clemson University, SC 29634-0978, U.S.A.\footnote{A member of the SARA Consortium}}

\author[0000-0002-6584-1703]{M. Ajello}
\affil{Department of Physics and Astronomy, Clemson University, SC 29634-0978, U.S.A.\footnote{A member of the SARA Consortium}}

\author[0000-0002-3433-4610]{A. Domínguez}
\affil{IPARCOS and Department of EMFTEL, Universidad Complutense de Madrid, E-28040 Madrid, Spain}

\author[0000-0001-5990-6243]{A. Rau}
\affil{Max-Planck-Institut für extraterrestrische Physik, Giessenbachstraße 1, D-85748 Garching, Germany}

\author{S. B. Cenko}
\affil{Astrophysics Science Division, NASA Goddard Space Flight Center, Mail Code 661, Greenbelt, MD 20771, USA}
\affil{Joint Space-Science Institute, University of Maryland, College Park, MD 20742, USA}

\author{J. Greiner}
\affil{Max-Planck-Institut für extraterrestrische Physik, Giessenbachstraße 1, D-85748 Garching, Germany}

\author[0000-0002-8028-0991]{D. H. Hartmann}
\affil{Department of Physics and Astronomy, Clemson University, SC 29634-0978, U.S.A.\footnote{A member of the SARA Consortium}}
\affil{Southeastern Association for Research in Astronomy (SARA), USA}

\author{A. Circiello}
\affil{Department of Physics and Astronomy, Clemson University, SC 29634-0978, U.S.A.\footnote{A member of the SARA Consortium}}

\author{I. Cox}
\affil{Department of Physics and Astronomy, Clemson University, SC 29634-0978, U.S.A.\footnote{A member of the SARA Consortium}}

\author[0000-0001-9427-2944]{S. Joffre}
\affil{Department of Physics and Astronomy, Clemson University, SC 29634-0978, U.S.A.\footnote{A member of the SARA Consortium}}

% \author{C. Karwin}
% \affil{Department of Physics and Astronomy, Clemson University, SC 29634-0978, U.S.A.\footnote{A member of the SARA Consortium}}

\author[0000-0002-8436-1254]{A. McDaniel}
\affil{Department of Physics and Astronomy, Clemson University, SC 29634-0978, U.S.A.\footnote{A member of the SARA Consortium}}

\author[0009-0004-5648-2405]{G. Rajguru}
\affil{Department of Physics and Astronomy, Clemson University, SC 29634-0978, U.S.A.\footnote{A member of the SARA Consortium}}

\author{R. Silver}
\affil{Department of Physics and Astronomy, Clemson University, SC 29634-0978, U.S.A.\footnote{A member of the SARA Consortium}}

\author{N. Torres-Albà}
\affil{Department of Astronomy, University of Virginia, VA 22904, U.S.A.}

\author{A. Webber}
\affil{Department of Physics and Astronomy, Clemson University, SC 29634-0978, U.S.A.\footnote{A member of the SARA Consortium}}

\begin{abstract}
\noindent Measuring spectroscopic redshifts for BL Lacertae (BL Lac) objects, a class of blazar, is challenging because their optical spectrum lacks, or has weak, emission lines ( equivalent width $\leqslant5\AA$). In this situation, alternative techniques are necessary for the estimation of distances to these sources. In this paper, we estimate the redshift by the photometric dropout technique for a sample of 64 blazars (59 BL Lacs and five blazar candidates of uncertain type). Two telescopes are utilized to observe the sample. The Ultraviolet/Optical Telescope (UVOT) on board {\it Swift} ({\it Swift}/UVOT) observes sources in $uvw2,\ uvm2,\ uvw1,\ u,\ b,\ v$ filters, while the ground-based telescopes SARA-CT/RM observed sources in $g',\ r,' \ i',\ z'$ filters. We fit the photometric data with the LePHARE package and report four new high-$z$ ($z>1.3$) BL Lacs at $2.03^{+0.07}_{-0.05}$, $1.84^{+0.10}_{-0.03}$, $2.04^{+0.16}_{-0.14}$, and $2.93^{+0.01}_{-0.04}$ as well as upper limits for 50 sources. This work increased the number of high-$z$ BL Lacs found by this method up to 23. The high-$z$ sources are discussed in the context of the cosmic gamma-ray horizon, blazar sequence, Fermi blazar divide, and masquerading BL Lacs.
\end{abstract}


\section{Introduction} \label{sec:intro}
As one of the most extreme environments in the universe, active galactic nuclei (AGNs) are characterized by intense gravitational forces, relativistic jets and luminous radiations \citep{urryUnifiedSchemesRadioLoud1995}. These are of great interest for studying particle acceleration mechanisms under extreme conditions unreachable in current laboratories. The supermassive black holes hosted at the centers of these galaxies \citep{fabianTestingAGNParadigm2008} accrete matter from the surrounding accretion disk, which powers the high luminosities of AGNs \citep{marconiLocalSupermassiveBlack2004}. 

The AGNs can be classified based on their orientation toward Earth \citep{urryUnifiedSchemesRadioLoud1995}. Blazars are a type of jetted AGN with the relativistic jet pointing along the line of sight towards us, viewed from an angle $\theta_{\nu}<1/\Gamma$ \citep{Blandford_1978, Marcotulli_2017}, where $\Gamma$ is the bulk Lorentz factor. They emit light that covers the entire electromagnetic spectrum from low-frequency radio to high-energy gamma rays, with high variability \citep{perlmanActiveGalacticNuclei2013}. The blazar spectral energy distribution (SED) is characterized by two peaks. The lower-energy peak is typically located from the infrared to the X-ray band, which is caused by the synchrotron radiation of relativistic electrons in the magnetic field. The higher-energy peak that goes from the X-ray to gamma-ray band is interpreted as the outcome of inverse-Compton scattering off the synchrotron photons \citep{maraschiBroadBandEnergy1994} or the circumnuclear photon field \citep{Dermer_1994_EC_IC}. The frequency of the synchrotron bump can also classify blazars as low synchrotron peaked (LSP) blazars with $\nu^{S}_{peak}\leqslant10^{14}$Hz, intermediate synchrotron peaked (ISP) blazars with $10^{14}\leqslant\nu^{S}_{peak}\leqslant10^{15}$Hz, and high synchrotron peaked (HSP) blazars with $\nu^{S}_{peak}\geqslant10^{15}$Hz \citep{abdoFermiLargeArea2010}. There is an additional classification by the emission lines in the optical spectroscopy of blazars \citep{padovaniConnectionXRayRadioselected1995}. If blazars show broad emission lines (equivalent width $\geqslant5\text{\AA}$) in their optical spectra, they are called Flat spectrum radio quasar (FSRQ). On the other hand, blazars that display no or weak emission lines (equivalent width $\leqslant5\text{\AA}$) are called BL Lacertae (BL Lac) \citep{maraschiBroadBandEnergy1994, Ajello_2013}. The featureless spectrum makes it challenging to measure redshifts via traditional optical spectroscopy. The spectroscopy measurement usually requires large telescopes ($>8$ meters) and a large amount of telescope time \citep{spectroscopy1, spectroscopy2}. As a result, nearly $38\%$ of the BL Lacs in the Fourth LAT AGN Catalog Data Release 3 (4LAC-DR3) lack redshift measurements \citep{Ajello_2020, ajello2022fourth}.

As an alternative method, the photometric redshift measurement has been implemented for sources like gamma-ray bursts (GRBs) \citep{tagliaferri2005grb, 2011Kruhler}. Photons from blazars interact with the neutral hydrogen along the line of sight, leaving two characteristic dropouts at rest wavelength bluewards of 912$\text{\AA}$ (Lyman limit) and 1216$\text{\AA}$ (Ly$\alpha$ forest). The position and the depth of the absorption features are dependent on the distance. Therefore, fitting the dropouts with SED templates can estimate the redshift. \cite{Rau2012} has implemented this technique using simultaneous multi-band photometric data from {\it Swift}/UVOT \citep{roming2005swift} and Gamma-ray burst Optical/NearInfrared Detector (GROND, \citealt{greiner2007grond}). The data are fitted to a list of SED templates to obtain the photometric redshift for a sample of 103 blazars from the Second LAT AGN Catalog (2LAC; \citealt{ackermannSecondCatalogActive2011}), which yielded six new high-$z$ ($z>1.3$) BL Lacs. The continued Photo-$z$ campaign has discovered 13 more new high-$z$ BL Lacs following the same method \citep{kaur2017, kaur2018, Rajagopal_2020, sheng2024revealing}.

The Extragalactic background light (EBL) is the integrated emission from star formation and supermassive black holes, and the reprocessed light due to the interstellar medium over all redshifts \citep{dominguezSpectralAnalysisFermiLAT2015, desai2019identifying}, expanding from infrared to optical \citep{franceschiniExtragalacticBackgroundLight2017}. Measuring the EBL directly is challenging due to the strong foreground and zodiacal light from the Milky Way \citep{moralejoMeasurementEBLCombined2017}. Nonetheless, blazars are appropriate objects for measuring the EBL indirectly because the attenuation produced by the interaction between the blazar gamma-ray photons and the EBL photons leaves a distinct fingerprint in the blazar SED. The attenuation in the SED can constrain the evolution of the EBL models with redshift \citep{Biteau_2015, Abdollahi_EBL_2018}. The higher the redshift, the stronger the attenuation; therefore, looking for high-$z$ sources is essential to probe the EBL \citep{Ackermann_2012}, motivating the search for high-$z$ BL Lacs. %Population studies of blazars also calls for more high-$z$ BL Lacs: the unified model, blazar sequence \citep{fossati1998unifying}, separates FSRQs and BL Lacs based on their spectral properties \citep{prandini2022blazar}. However, outliers are found by the photometric high-$z$ BL Lacs from the photo-z campaign \citep{sheng2024revealing}, indicating that the spectroscopic redshift surveys might bias the current population model because of the difficulty in the redshift measurement of BL Lac. Therefore, increasing the number of high-$z$ BL Lacs is essential to examine the unified blazar population model. 

This paper is a continuation of the previous campaigns. Photometric data in ten filters from near-infrared (NIR) to ultraviolet ($uvw2,\ \allowbreak uvm2,\ \allowbreak uvw1,\ \allowbreak u,\ \allowbreak b,\ \allowbreak v,\ \allowbreak g',\ \allowbreak r',\ \allowbreak i',\ \allowbreak z'$) are obtained from the Neil Gehrels Swift Observatory ({\it Swift}) \citep{gehrels2004swift}, 1.0-meter SARA-RM (Southeastern Association for Research in Astronomy at Roque de los Muchachos, La Palma, Spain), and 0.6-meter SARA-CT (Cerro Tololo, Chile) telescopes \citep{keel_remote_2016}. A flat cosmological model $\Lambda$CDM with $H_0=73\ \mathrm{km/s/Mpc}$, $\Omega_m=0.3$, $\Omega_{\Lambda}=0.7$ is assumed in this work. The paper is organized as: Section \ref{sec:observations} introduces the telescopes and observation techniques; Section \ref{sec:data analysis} explains the aperture photometry method; Section \ref{sec:sed_fitting} illustrates the SED fitting procedures to obtain the photometric redshifts; Section \ref{sec:redults} shows the high-$z$ BL Lacs found in this work; Section \ref{sec: discussions} is the application of high-$z$ BL Lacs; the last Section \ref{sec:summary} is a summary of the work presented.

\section{Observations} \label{sec:observations}

The studied sample in this work contains 64 blazars from the third Fermi Large Area Telescope source catalog (3FGL) \citep{collaboration_fermi_2015}, 4FGL-DR3 \citep{abdollahi2020fermi, abdollahi2022incremental}, and Third \emph{Fermi}-LAT Catalog of High-Energy Sources (3FHL, \citealt{ajello20173fhl}). It includes 59 BL Lacs and 5 blazar candidates of uncertain type (BCU) with E(B-V)$\leqslant0.3$ mag. The {\it Swift}/UVOT observations are obtained via Cycle 18\footnote{Proposal number: 1821042; PI: Yong Sheng} and 19\footnote{Proposal number: 1922139; PI: Yong Sheng} of the {\it Swift} investigator program or Targets of opportunities (ToOs). The mode \texttt{0x30ed} was used for the {\it Swift}/UVOT observation, resulting in a weighted time spent in each filter as $uvw2:uvm2:uvw1:u:b:v = 4:3:2:1:1:1$. The integrated time across the {\it Swift}/UVOT filters is $\sim$2000 seconds to ensure a good signal-to-noise (S/N).

The photometric data from the optical to NIR band are obtained by the ground observatory SARA-CT/RM. The two observatories are located in Spain and Chile, providing visibility from both hemispheres. However, the {\it Swift}/UVOT and SARA data are not always taken simultaneously, due to the availability of the SARA telescopes. The Sloan Digital Sky Survey ({\it SDSS}) filters ($g',r',i',z'$) are used for observations. Although the exposures might be split to avoid cosmic-ray and trailing effects, the integrated time over a filter is 15 to 40 minutes to guarantee good S/N. The targets are observed in sets, and each set contains the split exposures in the four {\it SDSS} filters. Table 1 presents the target list and the observation dates for our sample.

\section{Data analysis and calibrations} \label{sec:data analysis}

\subsection{{\it Swift} and SARA data analysis}
The {\it Swift}/UVOT and SARA data are analyzed via the Python package {\it photozpy}\footnote{https://github.com/Yong2Sheng/photozpy}. The automatic data analysis pipeline is built for the data reduction and aperture photometry. This package requires HEASoft\footnote{https://heasarc.gsfc.nasa.gov/docs/software/heasoft/} (v.6.31.1 for this paper) and the {\it Swift} Calibration Database (CALDB)\footnote{https://heasarc.gsfc.nasa.gov/docs/heasarc/caldb/} (v.1.0.2 for this paper) to be installed for the {\it Swift} data analysis. This is because the {\it swiftz} module is a wrapper for the \texttt{UVOTIMSUM}, \texttt{UVOTSOURCE}, and \texttt{fappend} commands for aligning and combining image extensions and running photometry. The {\it Swift} data analysis follows the standard pipeline provided by \cite{poole_photometric_2008}. The circular source region has a radius of 5${\arcsec}$ for all filters to match the calibration from counts to magnitudes. However, the circular background region has a radius varying from 20${\arcsec}$ to 30${\arcsec}$, depending on the source density of the field. Instrument response, bad pixel removal, sensitivity and dark corrections are handled by CALDB.

\begin{figure}[h]
    \centering
    \includegraphics[width=0.4\textwidth]{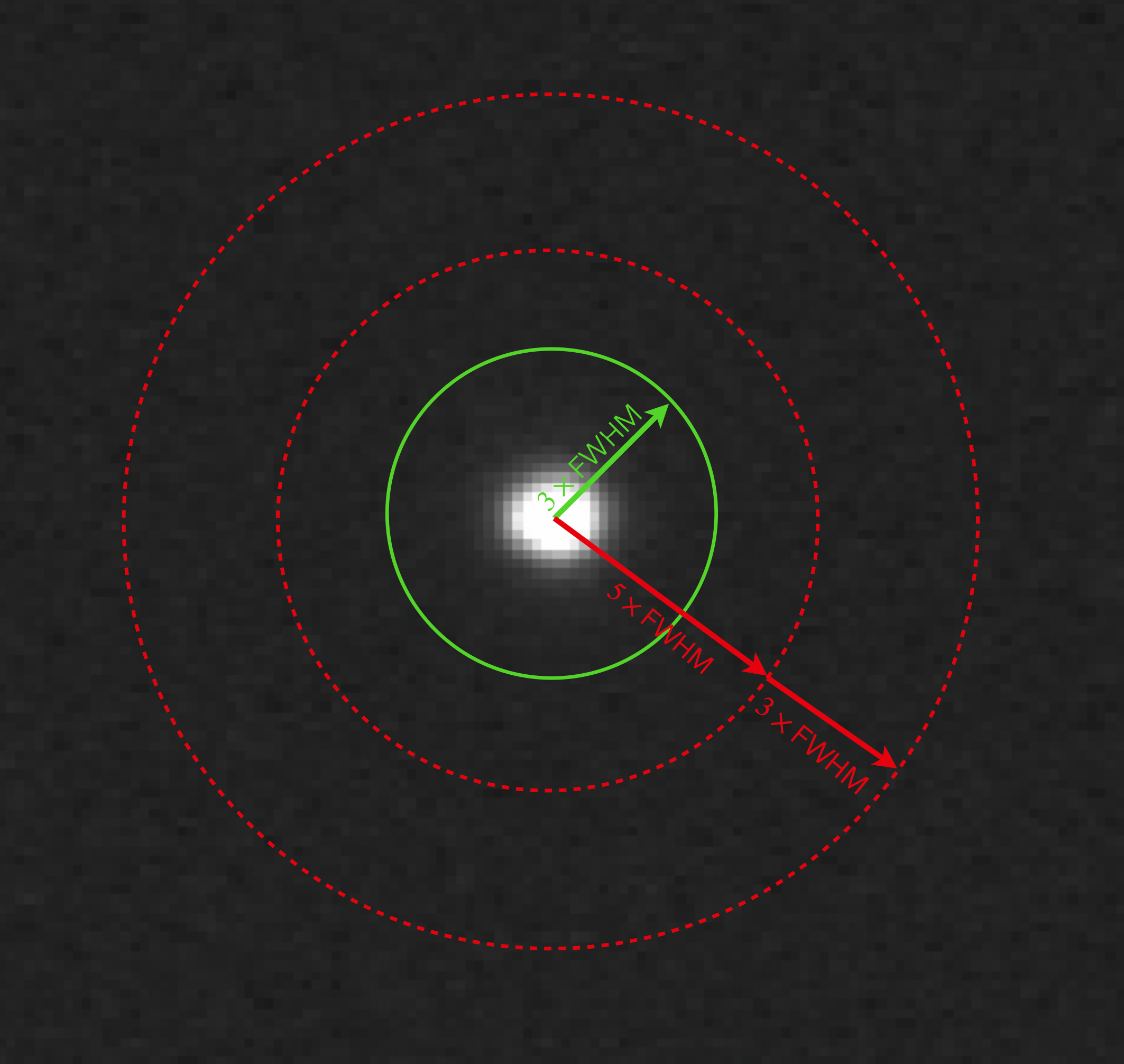}
    %\captionsetup{width=0.8\textwidth}
    \caption{The illustration of the aperture and background regions. The source aperture and background annulus for SARA data. The FWHM is measured in pixel scale. The aperture is centered at the target source with a radius of $3\times\text{FWHM}$. The background region is an annulus with an inner radius of $5\times\text{FWHM}$ and an outer radius of $8\times\text{FWHM}$, which ensures the background region area is $\sim4$ times the aperture one.}
    \label{fig:CCD_regions}
\end{figure}

The SARA CCD data was also analyzed using {\it photozpy}. The data reduction is handled by the {\it ccdproc} package \citep{matt_craig_2017_1069648}. The bias and dark frames remove the electronic and thermal noise. The flat frames correct pixel-to-pixel sensitivity variation of the CCD and the distortions in the optical path. 
Cosmic rays are removed by {\it ccdproc} prior to the image alignment by {\it astroalign} package \citep{BEROIZ2020100384} to avoid false source detection.
Plate solving is completed by {\it Astrometry.Net} \citep{hoggAutomatedAstrometryInvited2008, langASTROMETRYNETBLINDASTROMETRIC2010}.
The {\it photozpy} also provides source centroiding and full width at half maximum (FWHM) measurement. The FWHM is measured for all the images as the metric for the source and background radius for each night. Figure \ref{fig:CCD_regions} shows the radius selection for the source and background regions. The {\it photutils} package \citep{larry_bradley_2024_10967176} performs the aperture photometry, producing instrumental magnitudes. We use standard stars \citep{smith2006southern,landolt_ubvri_2009, albareti_13th_2017} covering both hemispheres to calibrate the instrumental magnitudes. Galactic extinction is applied for both SARA and {\it Swift}/UVOT magnitudes \citep{kataoka_multiwavelength_2008}.

\subsection{Variability correction and cross-calibration}

Both {\it Swift}/UVOT and SARA-CT/RM telescopes observed target sources in sequence in filters. Therefore, we considered variability, usually spanning from minutes to days \citep{racine1970photometry,carini1991timescales,urry1993multiwavelength,variability2022statistical}, during the SED fitting. We included an error of 0.1 magnitudes caused by the blazar variability in the systematic uncertainty. This error is based on the measurement in the $uvw2$ filter in \cite{Rau2012} and is implemented in the photo-$z$ technique.

\cite{Rau2012} also implemented a cross-calibration between the two instruments under the assumption that the SED of the targets remains the same. \cite{Rajagopal_2020} fitted a powerlaw SED with a quadratic function using color terms $g'-r'$ and $b-g'$ for the cross-calibration formula:
\begin{equation} \label{eq1}
b - g' = 0.26(g' - r') + 0.02(g' - r')^2.
\end{equation}
We calculated the offset in the $b$ filter and applied it to all the {\it Swift}/UVOT filters. Table \ref{tab:mags} is a summary of the cross-calibrated photometry.

\section{SED Fitting} \label{sec:sed_fitting}

LePHARE v2.2 \citep{arnouts1999measuring,ilbert2006accurate} is a package for photometric redshift estimation, which fits the photometric data to SED templates with known redshift. It uses $\chi^2$ statistics to check how well the data fits the templates and determines the best-fit SED model. \cite{Rau2012, Rajagopal_2020} assessed the robustness of the photometric redshift estimation using LePHARE simulations, testing power-law indices from 0 to 3, for redshifts ranging from 0 to 3, and for reddening E(B-V) up to 0.3. We use three different SED model templates for the sample. The first is a power-law library with 60 templates with power-law indices ranging from 0 to 3 ($F_{\lambda} \propto \lambda^{-\beta}, \beta \in  [0, 3]$) . The second library is a mix of normal galaxies, AGNs, and AGN/galaxy hybrids intended to fit sources dominated by the nucleus or the host galaxy \citep{salvato2008photometric,salvato2011dissecting}. The last library is a set of stellar templates to avoid false photometric redshift measurements \citep{bohlin1995white,pickles1998stellar,chabrier2000evolutionary}.

Only sources with $E(B-V)\leqslant0.3$ will have reliable photometric redshift measurements within$\left|\Delta z /\left(1+z_{\operatorname{sim}}\right)\right|<0.15$ for $z>1.3$, where $z_{sim}$ is the input redshift for the simulation \citep{Rajagopal_2020}. To ensure the fitted photometric redshift is within $0.1(1+z_{\text{phot}})$ of the best-fit value, we only report high-$z$ sources with a probability density of $P_z > 90\% $. We report the redshift upper limits for sources that do not meet the criteria but have $P_z < 90\% $ and $\chi^2 \leqslant30$. No redshift is reported for sources with $E(B-V)>0.3$ since it will produce an unreliable redshift \citep{Rajagopal_2020}.

%\clearpage
\startlongtable
\begin{deluxetable*}{llcccllc}%[!ht]
	\tablecolumns{7}
	\tablecaption{\label{tab:obs} $\emph{Swift}/$UVOT and SARA Observations along with visual extinction values. $A_{V}=3.1\times$E(B$-$V)}
	\tablewidth{0pt}
	\tabletypesize{\small}
	\setlength{\tabcolsep}{0.07in} 
	\tablehead{
		\colhead{{\it Fermi}}   & \colhead{$\emph{$\emph{Counterpart}$}$} & \colhead{Class} & \colhead{RA J2000$^a$} &	\colhead{Dec J2000$^a$} &	\colhead{$\emph{Swift}$/UVOT Date$^{b}$} & 	\colhead{SARA Date$^{b}$} & 	\colhead{$A_{V}$} \\
		\colhead{ (Name)}   & \colhead{(Name)} & \colhead{(Type)} & \colhead{(hh:mm:ss)} &	\colhead{($^\circ: ^\prime: ''$)} &	\colhead{(UT)} & 	\colhead{(UT)} & 	\colhead{(mag)}
    	}
\startdata
4FGL J0019.6$+$2022 & PKS 0017$+$200 & bll & 00:19:37.85 & $+$20:21:45.64 & 2022 Oct. 17 & 2022 Oct.16 & 0.1617\\ 
4FGL J0023.9$+$1603 & 87GB 002122.5$+$154553 & bll & 00:24:01.28 & $+$16:02:33.95 & 2023 Sep. 20 & 2023 Sep. 08 & 0.1758\\ 
4FGL J0026.6$-$4600 & 1RXS J002636.3$-$460101 & bll & 00:26:35.40 & $-$46:01:10.92 & 2023 Aug. 09 & 2023 Nov. 19 & 0.0269\\ 
3FHL J0121.8$+$3808 & ILT J012205.88$+$380445.2$^d$ & bll$^d$ & 01:22:06.02$^d$ & $+$38:04:45.41$^d$ & 2022 Nov. 19 & 2022 Nov. 06 & 0.1231\\ 
4FGL J0125.3$-$2548 & PKS 0122$-$260 & bll & 01:25:18.84 & $-$25:49:04.39 & 2022 Sep. 10 & 2022 Sep. 28 & 0.0401\\ 
4FGL J0138.0$+$2247 & GB6 J0138$+$2248 & bll & 01:38:01.14 & $+$22:48:08.57 & 2022 Nov. 20 & 2022 Nov. 11 & 0.3212\\ 
4FGL J0139.0$+$2601 & WISE J013859.14$+$260015.7 & bll & 01:38:59.15 & $+$26:00:15.71 & 2022 Sep. 20 & 2022 Oct.16 & 0.2535\\ 
4FGL J0144.6$+$2705 & TXS 0141$+$268 & bll & 01:44:33.55 & $+$27:05:03.12 & 2022 Dec. 06 & 2022 Nov. 27 & 0.1899\\ 
4FGL J0156.9$-$5301 & 1RXS J015658.6$-$530208 & bll & 01:56:58.00 & $-$53:01:59.98 & 2022 Aug. 24 & 2022 Sep. 28 & 0.067\\ 
4FGL J0237.3$+$2000 & NVSS J023720$+$200033 & bll & 02:37:19.77 & $+$20:00:31.76 & 2022 Sep. 30 & 2022 Oct.16 & 0.3625\\ 
4FGL J0338.5$+$1302 & RX J0338.4$+$1302 & bll & 03:38:29.30 & $+$13:02:15.45 & 2022 Dec. 31 & 2022 Nov. 30 & 0.9293\\ 
4FGL J0344.4$+$3432 & 1RXS J034424.5$+$343016 & bcu & 03:44:24.96 & $+$34:30:18.01 & 2023 Oct. 15 & 2023 Oct. 14 & 0.8228\\ 
4FGL J0406.0$-$5407 & SUMSS J040608$-$540445 & bll & 04:06:08.66 & $-$54:04:49.89 & 2023 Nov. 29 & 2023 Nov. 19 & 0.0262\\ 
4FGL J0439.8$-$1859 & 1SXPS J043949.5$-$190102 & bll & 04:39:49.73 & $-$19:01:01.42 & 2023 Jan. 22 & 2023 Jan. 16 & 0.1186\\ 
4FGL J0500.6$-$4911 & PMN J0500$-$4912 & bll & 05:00:38.80 & $-$49:12:16.59 & 2023 Jan. 16 & 2023 Jan. 12 & 0.029\\ 
4FGL J0506.9$+$0323 & NVSS J050650$+$032401 & bll & 05:06:50.15 & $+$03:23:58.76 & 2023 Oct. 18 & 2023 Oct. 14 & 0.1787\\ 
4FGL J0551.0$-$1622 & PMN J0550$-$1621 & bcu & 05:50:51.26 & $-$16:21:49.96 & 2024 Jan. 07 & 2024 Jan. 03 & 0.2194\\ 
4FGL J0558.8$-$7459 & PKS 0600$-$749 & bll & 05:58:46.04 & $-$74:59:05.21 & 2024 Jan. 01 & 2023 Dec. 26 & 0.3405\\ 
4FGL J0700.5$-$6610 & PKS 0700$-$661 & bll & 07:00:31.25 & $-$66:10:45.23 & 2024 Feb. 15 & 2024 Feb. 12 & 0.3043\\ 
4FGL J0746.6$-$4754 & PMN J0746$-$4755 & bll & 07:46:42.30 & $-$47:54:55.28 & 2024 Feb. 21 & 2024 Feb. 15 & 0.5356\\ 
4FGL J0816.9$+$2050 & SDSS J081649.78$+$205106.4 & bll & 08:16:49.78 & $+$20:51:06.44 & 2021 Mar. 15 & 2023 Apr. 18 & 0.1035\\ 
3FGL J0851.8$+$5531 & SDSS J085135.93$+$552834.5 & bll & 08:51:35.93 & $+$55:28:34.43 & 2021 Feb. 17 & 2022 Nov. 30 & 0.0983\\ 
4FGL J0854.3$+$4408 & B3 0850$+$443 & bll & 08:54:09.89 & $+$44:08:30.30 & 2021 Feb. 07 & 2022 Nov. 27 & 0.0704\\ 
4FGL J0859.4$+$6218 & 1RXS J085930.5$+$621737 & bll & 08:59:30.71 & $+$62:17:30.48 & 2021 Feb. 05 & 2022 Nov. 30 & 0.1657\\ 
4FGL J0910.6$+$3329 & Ton 1015 & bll & 09:10:37.03 & $+$33:29:24.43 & 2021 Jan. 27 & 2022 Nov. 27 & 0.0586\\ 
4FGL J0910.6$+$3329 & TXS 0956$-$244 & bll & 09:58:20.47 & $-$24:43:59.64 & 2024 Apr. 14 & 2024 Apr. 06 & 0.1409\\ 
4FGL J1019.7$+$6321 & GB6 J1019$+$6319 & bll & 10:19:50.87 & $+$63:20:01.62 & 2021 Oct. 27 & 2022 Nov. 30 & 0.0221\\ 
4FGL J1059.2$-$1134 & PKS B1056$-$113 & bll & 10:59:12.42 & $-$11:34:22.78 & 2023 Feb. 24 & 2023 Mar. 02 & 0.073\\ 
4FGL J1125.1$-$2101 & PMN J1125$-$2100 & bll & 11:25:08.62 & $-$21:01:06.00 & 2023 Apr. 11 & 2023 May 23 & 0.1872\\ 
4FGL J1248.3$+$5820 & PG 1246$+$586 & bll & 12:48:18.78 & $+$58:20:28.72 & 2021 Apr. 27 & 2023 Jun. 09 & 0.0295\\ 
4FGL J1259.8$-$3749 & NVSS J125949$-$374856 & bll & 12:59:49.80 & $-$37:48:58.16 & 2023 Feb. 12 & 2023 Jul. 07 & 0.1466\\ 
4FGL J1304.9$-$2107 & PKS B1302$-$208 & bcu & 13:04:59.07 & $-$21:06:42.46 & 2024 May. 11 & 2024 Apr. 06 & 0.3268\\ 
4FGL J1307.6$-$4259 & 1RXS J130737.8$-$425940 & bll & 13:07:37.98 & $-$42:59:38.99 & 2023 Feb. 08 & 2023 Mar. 02 & 0.2901\\ 
4FGL J1427.6$-$3305 & PKS 1424$-$328 & bll & 14:27:41.36 & $-$33:05:31.51 & 2023 Mar. 06 & 2023 May 23 & 0.1715\\ 
4FGL J1516.8$+$3651 & MG2 J151646$+$3650 & bll & 15:16:49.26 & $+$36:50:22.85 & 2022 Jun. 14 & 2023 Jun. 09 & 0.0511\\ 
4FGL J1612.4$-$3100 & NVSS J161219$-$305937 & bll & 16:12:20.00 & $-$30:59:38.67 & 2023 Apr. 20 & 2023 May 23 & 0.6209\\ 
4FGL J1624.6$+$5651 & SBS 1623$+$569 & bll & 16:24:32.18 & $+$56:52:27.98 & 2021 Jun. 28 & 2023 Apr. 18 & 0.0235\\ 
4FGL J1707.1$-$1931 & PMN J1706$-$1932 & bll & 17:06:58.69 & $-$19:31:51.81 & 2023 Apr. 26 & 2024 Jul. 17 & 0.7936\\ 
4FGL J1754.2$+$3212 & RX J1754.1$+$3212 & bll & 17:54:11.80 & $+$32:12:23.04 & 2022 Jul. 05 & 2022 Aug. 08 & 0.1127\\ 
4FGL J1823.5$+$6858 & 7C 1823$+$6856 & bll & 18:23:32.85 & $+$68:57:52.60 & 2022 Jul. 28 & 2022 Oct.16 & 0.1759\\ 
3FGL J1829.4$+$5402 & 1RXS J182925.7$+$540255 & bll & 18:29:24.28 & $+$54:02:59.75 & 2023 Jun. 14 & 2023 Jun. 09 & 0.0896\\ 
4FGL J1836.4$+$3137 & RX J1836.2$+$3136 & bll & 18:36:21.24 & $+$31:36:26.75 & 2022 Jun. 22 & 2022 Aug. 08 & 0.2757\\ 
4FGL J1838.8$+$4802 & GB6 J1838$+$4802 & bll & 18:38:49.15 & $+$48:02:34.26 & 2023 Jun. 08 & 2023 Jun. 09 & 0.1444\\ 
3FGL J1926.8$+$6154 & 1RXS J192649.5$+$615445 & bll & 19:26:49.89 & $+$61:54:42.34 & 2023 Jun. 29 & 2023 Jun. 21 & 0.1563\\ 
4FGL J1946.0$-$3112 & PKS 1942$-$313 & bll & 19:45:59.36 & $-$31:11:38.36 & 2023 May. 29 & 2023 May 23 & 0.3454\\ 
4FGL J1959.7$-$4725 & 1RXS J195945.8$-$472531 & bll & 19:59:45.67 & $-$47:25:19.35 & 2023 Jun. 01 & 2024 Jul. 17 & 0.1438\\ 
4FGL J2030.9$+$1935 & RX J2030.8$+$1935 & bll & 20:30:57.13 & $+$19:36:12.93 & 2023 Jun. 24 & 2023 Jun. 18 & 0.2461\\ 
4FGL J2126.1$-$3922 & PMN J2126$-$3921 & bll & 21:26:25.20 & $-$39:21:22.27 & 2022 Aug. 30 & 2022 Sep. 28 & 0.091\\ 
4FGL J2144.2$+$3132 & MG3 J214415$+$3132 & bll & 21:44:15.21 & $+$31:33:39.12 & 2023 Sep. 21 & 2023 Oct. 14 & 0.3884\\ 
4FGL J2149.6$+$0323 & PKS B2147$+$031 & bll & 21:49:41.87 & $+$03:22:51.43 & 2023 Aug. 01 & 2023 Aug. 18 & 0.2066\\ 
4FGL J2243.7$-$1231 & RBS 1888 & bll & 22:43:40.50 & $-$12:31:01.31 & 2021 Aug. 08 & 2023 Jul. 07 & 0.1421\\ 
4FGL J2243.9$+$2021 & RGB J2243$+$203 & bll & 22:43:54.74 & $+$20:21:03.78 & 2023 Aug. 26 & 2023 Sep. 08 & 0.1261\\ 
4FGL J2325.6$+$1644 & NVSS J232538$+$164641 & bll & 23:25:38.12 & $+$16:46:42.71 & 2021 Jul. 26 & 2022 Aug. 09 & 0.0974\\ 
4FGL J2352.0$+$1750 & CLASS J2352$+$1749 & bll & 23:52:05.84 & $+$17:49:13.75 & 2023 Sep. 27 & 2023 Sep. 24 & 0.118\\ 
3FHL J2358.4$-$1808 & NVSS 235836$-$180718$^d$ & bll$^d$ & 23:58:36.78$^d$ & $-$18:07:18.49$^d$ & 2022 May. 11 & 2022 Sep. 28 & 0.068\\ 
3FGL J0707.2$+$6101 & TXS 0702$+$612 & bcu & 07:07:00.57 & $+$61:10:10.56 & 2022 Nov. 23 & 2022 Nov. 27 & 0.2177\\ 
4FGL J0800.9$+$4401 & B3 0757$+$441 & bll & 08:01:08.28 & $+$44:01:10.16 & 2020 Nov. 30 & 2022 Nov. 27 & 0.1123\\ 
3FGL J0103.4$+$5336 & 1RXS J010325.9$+$533721 & bll & 01:03:25.96 & $+$53:37:13.30 & 2020 Jul. 27 & 2022 Aug. 15 & 0.9292\\ 
3FGL J1138.2$+$4905 & GB6 J1138$+$4858 & bcu$^{c}$ & 11:38:02.07 & $+$48:58:56.81 & 2023 Jan. 18 & 2023 Apr. 18 & 0.0466\\ 
3FGL J1323.9$+$1405 & RX J1323.9$+$1406 & bll & 13:23:58.49 & $+$14:06:01.08 & 2023 Jun. 01 & 2023 Jun. 21 & 0.0682\\ 
3FGL J1354.5$+$3705 & FIRST J135426.6$+$370654 & bll & 13:54:26.69 & $+$37:06:54.58 & 2023 Jun. 09 & 2023 Jun. 21 & 0.0297\\ 
3FGL J1500.6$+$4750 & BZB J1500$+$4751 & bll & 15:00:48.65 & $+$47:51:15.54 & 2023 Jun. 20 & 2023 May 29 & 0.0722\\ 
3FGL J1649.4$+$5238 & 87GB 164812.2$+$524023 & bll & 16:49:24.99 & $+$52:35:15.00 & 2023 Jun. 30 & 2023 Jul. 22 & 0.1482\\ 
3FGL J1702.6$+$3116 & RX J1702.6$+$3115 & bll & 17:02:38.80 & $+$31:15:46.66 & 2023 Jul. 10 & 2023 Sep. 08 & 0.0903\\ 
\enddata
    \end{deluxetable*}
\begin{footnotesize}
$^a$ The optical coordinates are from the 3FGL and 4FGL catalogs, which are used for \emph{Swift}/UVOT and SARA observations.

$^b$ The beginning date of the \emph{Swift}/UVOT and SARA observations.

$^{c}$ The source was a BCU when selected. It is classified as a FSRQ in the 4LAC catalog.
%\textbf{$^{c}$ The source was a BCU when selected. It is classified as a FSQR and the redshift is 0.354 as reported in the 4LAC catalog and \cite{marchesini2019optical}, which is compatible with the upper limit found in this paper.}

$^{d}$ The coordinates, counterparts and classes are determined by the method implemented in \cite{Joffre_2022, stroh2013swift,paiano2017optical,silver2020identifying,Kerby_2021}

\end{footnotesize}  
%=====================================================================================================

\begin{center}
\begin{longrotatetable}
%\begin{deluxetable*}{lllllllllll}
\begin{deluxetable*}{lllllllllll}
	\tablecolumns{11}
	\tablecaption{\label{tab:mags} $\emph{Swift}$/UVOT and SARA photometry (AB magnitudes corrected for extinction)}
	\tablewidth{0pt}
	\tabletypesize{\footnotesize}
	\setlength{\tabcolsep}{0.03in} 
	\tablehead{
\colhead{{\it Fermi} name}   & \colhead{$g^\prime$} & \colhead{$r^\prime$} &	\colhead{$i^\prime$} &	\colhead{$z^\prime$} & 	\colhead{$uvw2$} & 	\colhead{$uvm2$} & 	\colhead{$uvw1$} & 	\colhead{$u$} & \colhead{$b$} & \colhead{$v$}}
	\startdata
4FGL J0019.6$+$2022 & 20.22$\pm$0.12 & 19.43$\pm$0.11 & 18.99$\pm$0.15 & 19.06$\pm$0.46 & $>$22.59 & $>$21.9 & $>$21.81 & $>$21.17 & $>$20.44 & $>$19.62\\
4FGL J0023.9$+$1603 & 20.42$\pm$0.21 & 19.82$\pm$0.2 & 19.24$\pm$0.23 & $>$19.27 & $>$22.8 & $>$22.32 & $>$21.98 & $>$21.32 & $>$20.58 & $>$19.77\\
4FGL J0026.6$-$4600 & 18.13$\pm$0.08 & 17.26$\pm$0.04 & 18.38$\pm$0.21 & 17.13$\pm$0.18 & 19.11$\pm$0.07 & 18.87$\pm$0.08 & 18.82$\pm$0.07 & 18.5$\pm$0.07 & 18.38$\pm$0.09 & 18.09$\pm$0.13\\
3FHL J0121.8$+$3808 & 17.57$\pm$0.06 & 17.55$\pm$0.07 & 17.13$\pm$0.07 & 17.17$\pm$0.14 & 17.94$\pm$0.05 & 18.03$\pm$0.07 & 17.88$\pm$0.07 & 17.65$\pm$0.06 & 17.57$\pm$0.07 & 17.42$\pm$0.12\\
4FGL J0125.3$-$2548 & 19.96$\pm$0.19 & 19.68$\pm$0.21 & 19.45$\pm$0.32 & 17.64$\pm$0.2 & 21.95$\pm$0.23 & 21.46$\pm$0.28 & 21.8$\pm$0.36 & 20.31$\pm$0.2 & 20.03$\pm$0.27 & $>$19.51\\
4FGL J0138.0$+$2247 & 18.08$\pm$0.07 & 17.72$\pm$0.06 & $>$20.22 & 17.43$\pm$0.13 & 19.58$\pm$0.09 & 19.62$\pm$0.13 & 19.26$\pm$0.11 & 18.78$\pm$0.11 & 18.18$\pm$0.11 & 18.2$\pm$0.2\\
4FGL J0139.0$+$2601 & 17.65$\pm$0.02 & 17.11$\pm$0.02 & 16.84$\pm$0.03 & 16.64$\pm$0.05 & 19.08$\pm$0.07 & 18.99$\pm$0.09 & 18.49$\pm$0.08 & 18.15$\pm$0.07 & 17.79$\pm$0.07 & 17.55$\pm$0.1\\
4FGL J0144.6$+$2705 & 19.78$\pm$0.09 & 19.45$\pm$0.12 & 18.78$\pm$0.11 & 18.13$\pm$0.19 & $>$21.9 & $>$21.56 & $>$22.01 & $>$21.03 & $>$19.87 & $>$19.16\\
4FGL J0156.9$-$5301 & 17.96$\pm$0.03 & 17.51$\pm$0.03 & 17.33$\pm$0.06 & 16.74$\pm$0.11 & 18.74$\pm$0.07 & 18.76$\pm$0.09 & 18.54$\pm$0.06 & 18.39$\pm$0.08 & 18.08$\pm$0.09 & 17.82$\pm$0.14\\
4FGL J0237.3$+$2000 & 18.81$\pm$0.08 & 18.37$\pm$0.07 & 18.24$\pm$0.1 & 18.28$\pm$0.26 & 19.99$\pm$0.19 & 19.84$\pm$0.25 & 19.47$\pm$0.21 & 19.02$\pm$0.22 & 18.93$\pm$0.36 & $>$18.22\\
4FGL J0338.5$+$1302 & 17.56$\pm$0.08 & 17.46$\pm$0.06 & 17.34$\pm$0.08 & 17.47$\pm$0.21 & 18.35$\pm$0.16 & 18.73$\pm$0.22 & 18.42$\pm$0.22 & 17.68$\pm$0.15 & 17.58$\pm$0.19 & 17.15$\pm$0.22\\
4FGL J0344.4$+$3432 & 18.47$\pm$0.09 & 18.12$\pm$0.09 & 17.61$\pm$0.09 & 16.85$\pm$0.12 & $>$19.5 & $>$19.4 & $>$19.2 & $>$19.07 & $>$18.57 & 17.82$\pm$0.31\\
4FGL J0406.0$-$5407 & 18.9$\pm$0.11 & 18.09$\pm$0.07 & 17.73$\pm$0.1 & 18.78$\pm$0.84 & 20.39$\pm$0.09 & 20.19$\pm$0.13 & 19.97$\pm$0.11 & 19.34$\pm$0.1 & 19.12$\pm$0.12 & 18.95$\pm$0.21\\
4FGL J0439.8$-$1859 & 17.7$\pm$0.04 & 17.47$\pm$0.04 & 17.41$\pm$0.08 & 16.57$\pm$0.12 & 18.77$\pm$0.06 & 18.84$\pm$0.08 & 18.45$\pm$0.08 & 18.08$\pm$0.07 & 17.76$\pm$0.07 & 17.64$\pm$0.11\\
4FGL J0500.6$-$4911 & 18.22$\pm$0.04 & 17.72$\pm$0.03 & 16.92$\pm$0.03 & 17.98$\pm$0.3 & 19.44$\pm$0.09 & 19.23$\pm$0.11 & 18.9$\pm$0.1 & 18.71$\pm$0.12 & 18.35$\pm$0.14 & 17.72$\pm$0.17\\
4FGL J0506.9$+$0323 & 19.08$\pm$0.08 & 18.72$\pm$0.1 & 18.64$\pm$0.16 & $>$19.05 & 20.7$\pm$0.21 & 20.5$\pm$0.28 & 20.31$\pm$0.27 & 19.69$\pm$0.25 & 19.18$\pm$0.3 & 18.52$\pm$0.33\\
4FGL J0551.0$-$1622 & $>$20.5 & $>$20.33 & $>$19.77 & $>$18.44 & $>$22.38 & $>$22.01 & $>$21.58 & $>$21.23 & $>$20.55 & $>$19.79\\
4FGL J0558.8$-$7459 & 19.18$\pm$0.97 & 18.28$\pm$0.32 & 18.48$\pm$0.42 & 18.48$\pm$0.86 & 21.55$\pm$0.33 & $>$21.17 & 20.7$\pm$0.26 & 19.87$\pm$0.2 & 19.44$\pm$0.22 & 19.27$\pm$0.36\\
4FGL J0700.5$-$6610 & 16.15$\pm$0.01 & 15.78$\pm$0.01 & 15.51$\pm$0.02 & 15.34$\pm$0.04 & 17.49$\pm$0.05 & 17.51$\pm$0.06 & 17.12$\pm$0.06 & 16.65$\pm$0.05 & 16.25$\pm$0.04 & 16.1$\pm$0.06\\
4FGL J0746.6$-$4754 & 16.27$\pm$0.03 & 16.0$\pm$0.02 & 15.83$\pm$0.03 & 15.7$\pm$0.07 & 17.28$\pm$0.07 & 17.2$\pm$0.09 & 16.91$\pm$0.08 & 16.59$\pm$0.05 & 16.35$\pm$0.05 & 16.11$\pm$0.08\\
4FGL J0816.9$+$2050 & 18.24$\pm$0.03 & $>$21.18 & 18.83$\pm$0.17 & $>$19.46 & 19.22$\pm$0.16 & 18.88$\pm$0.17 & 18.45$\pm$0.11 & 17.97$\pm$0.14 & 17.65$\pm$0.24 & 16.93$\pm$0.25\\
3FGL J0851.8$+$5531 & 18.01$\pm$0.02 & 17.56$\pm$0.02 & 17.22$\pm$0.04 & 17.04$\pm$0.1 & 19.21$\pm$0.08 & 18.98$\pm$0.09 & 18.63$\pm$0.09 & 18.3$\pm$0.08 & 18.14$\pm$0.1 & 17.49$\pm$0.12\\
4FGL J0854.3$+$4408 & 16.67$\pm$0.01 & 16.42$\pm$0.01 & 16.27$\pm$0.01 & 16.15$\pm$0.03 & 17.51$\pm$0.04 & 17.37$\pm$0.05 & 17.24$\pm$0.05 & 16.95$\pm$0.04 & 16.73$\pm$0.04 & 16.6$\pm$0.07\\
4FGL J0859.4$+$6218 & 19.54$\pm$0.07 & 19.13$\pm$0.08 & 18.94$\pm$0.14 & 18.68$\pm$0.33 & 20.64$\pm$0.09 & 20.43$\pm$0.11 & 20.33$\pm$0.11 & 20.0$\pm$0.12 & 19.65$\pm$0.14 & 19.34$\pm$0.22\\
4FGL J0910.6$+$3329 & 16.92$\pm$0.01 & 16.64$\pm$0.01 & 16.47$\pm$0.01 & 16.36$\pm$0.03 & 17.9$\pm$0.04 & 17.77$\pm$0.05 & 17.6$\pm$0.05 & 17.2$\pm$0.04 & 17.0$\pm$0.04 & 16.89$\pm$0.06\\
4FGL J0910.6$+$3329 & 20.71$\pm$0.7 & 20.08$\pm$0.7 & 19.5$\pm$0.71 & $>$18.44 & $>$22.56 & $>$22.15 & $>$21.59 & $>$21.45 & $>$20.88 & $>$20.15\\
4FGL J1019.7$+$6321 & 17.93$\pm$0.02 & 17.57$\pm$0.02 & 17.46$\pm$0.05 & 17.69$\pm$0.21 & 19.24$\pm$0.09 & 19.25$\pm$0.12 & 19.01$\pm$0.12 & 18.59$\pm$0.15 & 18.03$\pm$0.17 & 17.89$\pm$0.31\\
4FGL J1059.2$-$1134 & 17.44$\pm$0.06 & 17.06$\pm$0.03 & 16.66$\pm$0.04 & 16.14$\pm$0.07 & 18.86$\pm$0.05 & 18.69$\pm$0.07 & 18.46$\pm$0.06 & 18.5$\pm$0.05 & 17.54$\pm$0.05 & 17.29$\pm$0.07\\
4FGL J1125.1$-$2101 & 18.47$\pm$0.07 & 18.28$\pm$0.1 & 19.15$\pm$0.36 & $>$18.57 & 19.87$\pm$0.07 & 19.65$\pm$0.09 & 19.37$\pm$0.08 & 18.78$\pm$0.06 & 18.52$\pm$0.07 & 18.19$\pm$0.1\\
4FGL J1248.3$+$5820 & 15.78$\pm$0.0 & 15.4$\pm$0.0 & 15.16$\pm$0.0 & 14.93$\pm$0.01 & 16.91$\pm$0.04 & 16.73$\pm$0.04 & 16.6$\pm$0.04 & 16.16$\pm$0.04 & 15.88$\pm$0.04 & 15.63$\pm$0.04\\
4FGL J1259.8$-$3749 & $>$20.18 & 18.3$\pm$0.15 & 18.99$\pm$1.03 & $>$17.9 & 19.73$\pm$0.09 & 19.56$\pm$0.12 & 19.57$\pm$0.13 & 19.21$\pm$0.14 & 18.6$\pm$0.14 & 17.88$\pm$0.16\\
4FGL J1304.9$-$2107 & 20.45$\pm$0.72 & 19.88$\pm$0.65 & 19.45$\pm$0.69 & $>$18.49 & 22.34$\pm$0.35 & $>$21.91 & $>$21.57 & $>$21.26 & $>$20.6 & $>$19.9\\
4FGL J1307.6$-$4259 & 16.54$\pm$0.01 & 16.24$\pm$0.02 & 16.05$\pm$0.02 & 15.92$\pm$0.06 & 17.76$\pm$0.06 & 17.69$\pm$0.08 & 17.37$\pm$0.06 & 16.81$\pm$0.05 & 16.62$\pm$0.05 & 16.53$\pm$0.08\\
4FGL J1427.6$-$3305 & 19.22$\pm$0.15 & 17.89$\pm$0.07 & 17.62$\pm$0.08 & 17.02$\pm$0.1 & 21.36$\pm$0.13 & 21.44$\pm$0.2 & 21.04$\pm$0.18 & 20.31$\pm$0.15 & 19.6$\pm$0.14 & 19.17$\pm$0.19\\
4FGL J1516.8$+$3651 & 19.95$\pm$0.08 & 19.33$\pm$0.08 & 19.04$\pm$0.13 & 18.96$\pm$0.39 & $>$22.62 & $>$22.16 & 21.69$\pm$0.33 & 20.97$\pm$0.31 & 20.12$\pm$0.28 & $>$19.63\\
4FGL J1612.4$-$3100 & 18.25$\pm$0.12 & 17.63$\pm$0.09 & 17.36$\pm$0.09 & 17.59$\pm$0.22 & 19.69$\pm$0.25 & $>$18.88 & 19.57$\pm$0.29 & 18.64$\pm$0.18 & 18.42$\pm$0.24 & $>$17.7\\
4FGL J1624.6$+$5651 & 18.48$\pm$0.03 & 17.94$\pm$0.04 & 17.59$\pm$0.05 & 17.96$\pm$0.17 & 19.76$\pm$0.09 & 19.71$\pm$0.13 & 19.32$\pm$0.1 & 18.97$\pm$0.1 & 18.62$\pm$0.12 & 18.47$\pm$0.2\\
4FGL J1707.1$-$1931 & $>$19.32 & 18.1$\pm$0.2 & 17.75$\pm$0.18 & 17.77$\pm$0.39 & $>$20.93 & $>$20.68 & $>$20.55 & $>$20.27 & $>$19.67 & $>$19.05\\
4FGL J1754.2$+$3212 & 17.39$\pm$0.05 & 17.1$\pm$0.05 & 16.88$\pm$0.05 & 16.66$\pm$0.09 & 18.39$\pm$0.06 & 18.32$\pm$0.07 & 18.13$\pm$0.07 & 17.66$\pm$0.06 & 17.47$\pm$0.06 & 17.31$\pm$0.1\\
4FGL J1823.5$+$6858 & 19.41$\pm$0.06 & 18.99$\pm$0.07 & 18.57$\pm$0.11 & 18.7$\pm$0.35 & $>$22.36 & 21.82$\pm$0.33 & 21.4$\pm$0.3 & 20.02$\pm$0.16 & 19.52$\pm$0.18 & 18.9$\pm$0.21\\
3FGL J1829.4$+$5402 & 18.28$\pm$0.03 & 18.02$\pm$0.04 & 17.82$\pm$0.05 & 17.62$\pm$0.12 & 19.45$\pm$0.08 & 19.25$\pm$0.09 & 18.94$\pm$0.09 & 18.67$\pm$0.08 & 18.35$\pm$0.09 & 18.08$\pm$0.14\\
4FGL J1836.4$+$3137 & 17.95$\pm$0.1 & 17.23$\pm$0.06 & 16.89$\pm$0.06 & 16.93$\pm$0.12 & 19.62$\pm$0.12 & 19.61$\pm$0.16 & 18.98$\pm$0.12 & 18.7$\pm$0.12 & 18.15$\pm$0.13 & 17.56$\pm$0.15\\
4FGL J1838.8$+$4802 & 14.79$\pm$0.0 & 14.56$\pm$0.0 & 14.39$\pm$0.0 & 14.26$\pm$0.01 & 15.67$\pm$0.04 & 15.56$\pm$0.04 & 15.41$\pm$0.04 & 15.05$\pm$0.04 & 14.85$\pm$0.04 & 14.75$\pm$0.04\\
3FGL J1926.8$+$6154 & 17.31$\pm$0.01 & 17.09$\pm$0.01 & 16.86$\pm$0.02 & 16.8$\pm$0.06 & 18.0$\pm$0.05 & 17.87$\pm$0.08 & 17.78$\pm$0.06 & 17.42$\pm$0.05 & 17.37$\pm$0.06 & 17.24$\pm$0.09\\
4FGL J1946.0$-$3112 & 15.97$\pm$0.01 & 15.36$\pm$0.01 & 15.38$\pm$0.01 & 15.32$\pm$0.03 & 17.65$\pm$0.18 & 17.28$\pm$0.19 & 17.19$\pm$0.21 & 16.43$\pm$0.17 & 16.14$\pm$0.22 & 15.26$\pm$0.2\\
4FGL J1959.7$-$4725 & 16.47$\pm$0.02 & 16.11$\pm$0.02 & 15.88$\pm$0.03 & 15.69$\pm$0.08 & 17.44$\pm$0.05 & 17.36$\pm$0.06 & 17.16$\pm$0.06 & 16.7$\pm$0.04 & 16.56$\pm$0.05 & 16.5$\pm$0.08\\
4FGL J2030.9$+$1935 & 18.15$\pm$0.02 & 17.8$\pm$0.03 & 17.56$\pm$0.04 & 17.6$\pm$0.14 & 18.95$\pm$0.08 & 19.21$\pm$0.14 & 18.96$\pm$0.1 & 18.51$\pm$0.1 & 18.24$\pm$0.12 & 17.71$\pm$0.14\\
4FGL J2126.1$-$3922 & 19.98$\pm$0.21 & 19.7$\pm$0.24 & 19.27$\pm$0.3 & $>$18.6 & $>$22.36 & $>$21.86 & 21.38$\pm$0.32 & 20.63$\pm$0.31 & $>$20.05 & $>$19.19\\
4FGL J2144.2$+$3132 & 17.57$\pm$0.02 & 17.15$\pm$0.03 & 16.82$\pm$0.03 & 16.44$\pm$0.07 & 21.81$\pm$0.36 & 20.96$\pm$0.26 & 20.5$\pm$0.21 & 19.03$\pm$0.1 & 17.68$\pm$0.06 & 17.26$\pm$0.08\\
4FGL J2149.6$+$0323 & 18.1$\pm$0.03 & 17.65$\pm$0.03 & 17.35$\pm$0.04 & 17.07$\pm$0.1 & 19.61$\pm$0.12 & 19.25$\pm$0.15 & 19.0$\pm$0.13 & 18.47$\pm$0.12 & 18.23$\pm$0.16 & 17.94$\pm$0.23\\
4FGL J2243.7$-$1231 & $>$18.84 & $>$18.62 & $>$18.23 & $>$17.61 & 19.72$\pm$0.09 & 19.75$\pm$0.11 & 19.57$\pm$0.11 & 19.42$\pm$0.14 & 18.9$\pm$0.15 & 19.12$\pm$0.35\\
4FGL J2243.9$+$2021 & 16.05$\pm$0.0 & 15.78$\pm$0.01 & 15.59$\pm$0.01 & 15.44$\pm$0.02 & 17.06$\pm$0.04 & 16.88$\pm$0.05 & 16.71$\pm$0.04 & 16.36$\pm$0.04 & 16.13$\pm$0.04 & 15.92$\pm$0.05\\
4FGL J2325.6$+$1644 & 18.69$\pm$0.03 & 18.26$\pm$0.03 & 18.0$\pm$0.06 & 18.09$\pm$0.19 & 19.39$\pm$0.11 & 19.13$\pm$0.14 & 19.19$\pm$0.14 & 19.04$\pm$0.18 & 18.8$\pm$0.25 & $>$18.26\\
4FGL J2352.0$+$1750 & 18.05$\pm$0.04 & 17.71$\pm$0.04 & 17.64$\pm$0.06 & 17.82$\pm$0.2 & 19.39$\pm$0.09 & 19.24$\pm$0.13 & 18.93$\pm$0.11 & 18.47$\pm$0.1 & 18.15$\pm$0.13 & 17.71$\pm$0.18\\
3FHL J2358.4$-$1808 & 17.87$\pm$0.03 & 17.6$\pm$0.03 & 17.43$\pm$0.05 & 17.0$\pm$0.11 & 19.15$\pm$0.07 & 19.11$\pm$0.1 & ... & 18.41$\pm$0.09 & 17.94$\pm$0.1 & 18.13$\pm$0.24\\
3FGL J0707.2$+$6101 & 18.05$\pm$0.02 & 17.42$\pm$0.02 & 16.95$\pm$0.03 & 16.6$\pm$0.06 & 20.01$\pm$0.11 & 20.05$\pm$0.15 & 19.42$\pm$0.12 & 18.98$\pm$0.12 & 18.22$\pm$0.1 & 17.99$\pm$0.16\\
4FGL J0800.9$+$4401 & 19.09$\pm$0.05 & 18.57$\pm$0.05 & 18.4$\pm$0.1 & 18.59$\pm$0.36 & $>$21.58 & $>$21.11 & 20.48$\pm$0.29 & 19.83$\pm$0.28 & 19.23$\pm$0.32 & $>$18.52\\
3FGL J0103.4$+$5336 & 18.45$\pm$0.12 & 17.82$\pm$0.06 & 17.5$\pm$0.05 & 17.2$\pm$0.09 & 19.89$\pm$0.16 & 20.57$\pm$0.28 & 20.12$\pm$0.22 & 19.26$\pm$0.13 & 18.62$\pm$0.11 & 17.93$\pm$0.11\\
3FGL J1138.2$+$4905 & 20.88$\pm$0.29 & 20.22$\pm$0.31 & $>$20.54 & $>$19.56 & $>$23.24 & $>$22.73 & 22.57$\pm$0.36 & $>$21.84 & $>$21.05 & $>$20.0\\
3FGL J1323.9$+$1405 & 18.07$\pm$0.02 & 17.8$\pm$0.03 & 17.64$\pm$0.05 & 17.56$\pm$0.16 & 18.97$\pm$0.07 & 18.62$\pm$0.08 & 18.47$\pm$0.08 & 18.07$\pm$0.07 & 18.14$\pm$0.1 & 17.71$\pm$0.14\\
3FGL J1354.5$+$3705 & 17.99$\pm$0.02 & 17.48$\pm$0.02 & 17.07$\pm$0.03 & 16.64$\pm$0.06 & 19.46$\pm$0.06 & 19.13$\pm$0.07 & 19.06$\pm$0.08 & 18.49$\pm$0.06 & 18.13$\pm$0.07 & 17.83$\pm$0.1\\
3FGL J1500.6$+$4750 & 19.78$\pm$0.23 & 19.73$\pm$0.29 & 20.25$\pm$0.73 & $>$19.27 & 21.55$\pm$0.34 & $>$20.5 & 20.56$\pm$0.24 & 20.12$\pm$0.27 & $>$19.79 & 18.03$\pm$0.27\\
3FGL J1649.4$+$5238 & 17.54$\pm$0.01 & 17.12$\pm$0.02 & 16.67$\pm$0.02 & 16.26$\pm$0.05 & 18.94$\pm$0.08 & 18.81$\pm$0.09 & 18.61$\pm$0.09 & 18.0$\pm$0.08 & 17.65$\pm$0.08 & 17.18$\pm$0.11\\
3FGL J1702.6$+$3116 & 18.29$\pm$0.03 & 18.01$\pm$0.03 & 17.9$\pm$0.06 & 18.4$\pm$0.29 & 19.58$\pm$0.09 & 19.38$\pm$0.1 & 19.2$\pm$0.1 & 18.75$\pm$0.09 & 18.36$\pm$0.1 & 18.56$\pm$0.21\\
\enddata
\end{deluxetable*}
\end{longrotatetable}
\end{center}

%%%%%%%%%%%%%%%%%%%%%%%%%%%%%%%%%%%%%%%%%%%%%%%%%%%%%%%%%%%%%%%%%%%%%%%%%%%%%%%%%%%%%%%%%%%%

\startlongtable
%\centering
\begin{deluxetable*}{lccc|ccll|lcll}
\centering
	\tablecolumns{10}
	\tabletypesize{\small}
	%\rotate
	\tablewidth{0pt}
	%\tablenum{}
	\tablecaption{\label{tab:result}SED fitting}
	%\tablehead{}
	\setlength{\tabcolsep}{0.01in} 
	\tablehead{
		\colhead{3FGL or 3FHL Name}  & \colhead{$z_{\rm phot, best}^a$}  & \colhead{$z_{\rm spec,4LAC}^b$}& \colhead{$z_{\rm spec,{\it SDSS}}^b$} & \multicolumn{4}{c}{Power Law Template} & \multicolumn{4}{c}{Galaxies, AGNs, and AGN/galaxy hybrids} \\
		\cmidrule[\heavyrulewidth](lr){5-8}   \cmidrule[\heavyrulewidth](lr){9-12}
		\colhead{} & \colhead{} & \colhead{} & \colhead{} &\colhead{$z_{\rm phot}  ^c$} & \colhead{$\chi^{2\ d}$} & \colhead{$P_{\rm z}^e$} & \colhead{$\beta^f$} \hspace{0.4cm} & \colhead{$z_{\rm phot}  ^c$} & \colhead{$\chi^{2\ d}$} & \colhead{$P_{\rm z}^e$} & \colhead{Model}} 
    \startdata
	\multicolumn{10}{c}{\textbf{\small Sources with confirmed photometric redshifts}} \\
	\hline
4FGL J0506.9$+$0323 & $2.03^{+0.07}_{-0.05}$ & ... & ... & $2.03^{+0.07}_{-0.05}$ & 19.1 & 96.7 & 0.0 & $1.13^{+0.01}_{-0.01}$ & 15.4 & 79.5 & pl\_QSOH\_template\_norm.sed\\
4FGL J1125.1$-$2101 & $1.84^{+0.10}_{-0.03}$ & ... & ... & $1.84^{+0.10}_{-0.03}$ & 23.6 & 99.5 & 0.0 & $1.28^{+0.00}_{-0.03}$ & 22.5 & 93.7 & pl\_QSO\_DR2\_029\_t0.spec\\
4FGL J1823.5$+$6858 & $2.04^{+0.16}_{-0.14}$ & ... & $2.14^h$ & $2.04^{+0.16}_{-0.14}$ & 4.0 & 93.1 & 1.45 & $2.13^{+0.04}_{-0.06}$ & 10.6 & 93.4 & pl\_QSOH\_template\_norm.sed\\
4FGL J2144.2$+$3132 & $2.93^{+0.01}_{-0.04}$ & ... & ... & $2.93^{+0.01}_{-0.04}$ & 20.6 & 100.0 & 0.8 & $0.04^{+0.01}_{-0.01}$ & 12.2 & 98.8 & S0\_70\_QSO2\_30.sed\\
	\hline
	\multicolumn{10}{c}{\textbf{\small Sources with photometric redshift upper limits}}\\
	\hline
4FGL J0019.6$+$2022 & $<3.56$ & ... & 0.09 & $3.23^{+0.33}_{-0.49}$ & 0.6 & 75.9 & 1.45 & $0.20^{+0.06}_{-0.20}$ & 0.1 & 59.3 & S0\_90\_QSO2\_10.sed\\
4FGL J0023.9$+$1603 & $<3.55$ & ... & 0.77 & $3.14^{+0.41}_{-0.16}$ & 1.0 & 69.9 & 1.15 & $0.09^{+0.13}_{-0.05}$ & 0.0 & 29.9 & S0\_70\_QSO2\_30.sed\\
4FGL J0026.6$-$4600 & $...$ & ... & ... & $1.01^{+0.20}_{-1.01}$ & 38.9 & 33.5 & 0.85 & $0.00^{+0.01}_{-0.00}$ & 38.3 & 14.3 & pl\_I22491\_30\_TQSO1\_70.sed\\
3FHL J0121.8$+$3808 & $<0.99$ & ... & ... & $0.12^{+0.87}_{-0.12}$ & 4.7 & 22.1 & 0.5 & $0.16^{+0.40}_{-0.13}$ & 2.8 & 42.3 & pl\_TQSO1\_template\_norm.sed\\
4FGL J0125.3$-$2548 & $<1.37$ & ... & ... & $1.00^{+0.37}_{-1.00}$ & 17.3 & 27.3 & 1.95 & $0.00^{+0.07}_{-0.00}$ & 19.0 & 42.4 & Mrk231\_template\_norm.sed\\
4FGL J0138.0$+$2247 & $...$ & ... & 1.45 & $1.38^{+0.00}_{-0.01}$ & 220.9 & 98.6 & 3.0 & $...$ & ... & ... & ...\\
4FGL J0139.0$+$2601 & $<1.13$ & ... & 0.35 & $0.87^{+0.26}_{-0.87}$ & 4.2 & 32.7 & 1.5 & $0.00^{+0.01}_{-0.00}$ & 12.2 & 99.0 & I22491\_90\_TQSO1\_10.sed\\
4FGL J0144.6$+$2705 & $<2.85$ & ... & ... & $2.62^{+0.23}_{-0.00}$ & 4.6 & 81.5 & 2.0 & $0.06^{+0.04}_{-0.06}$ & 7.0 & 62.6 & S0\_60\_QSO2\_40.sed\\
4FGL J0156.9$-$5301 & $<0.89$ & ... & ... & $0.24^{+0.65}_{-0.24}$ & 9.8 & 27.2 & 1.1 & $0.00^{+0.30}_{-0.00}$ & 5.9 & 28.8 & I22491\_50\_TQSO1\_50.sed\\
4FGL J0237.3$+$2000 & $<1.46$ & ... & ... & $1.23^{+0.23}_{-1.23}$ & 2.0 & 33.4 & 1.0 & $1.24^{+0.12}_{-0.08}$ & 2.9 & 80.3 & pl\_QSOH\_template\_norm.sed\\
4FGL J0338.5$+$1302 & $<1.84$ & ... & ... & $1.73^{+0.11}_{-0.97}$ & 12.2 & 29.5 & 0.3 & $1.10^{+0.34}_{-0.11}$ & 11.0 & 65.9 & pl\_QSOH\_template\_norm.sed\\
4FGL J0344.4$+$3432 & $<2.26$ & ... & ... & $1.12^{+1.14}_{-1.12}$ & 6.3 & 17.6 & 2.0 & $1.47^{+0.01}_{-1.47}$ & 8.4 & 27.7 & Sey2\_template\_norm.sed\\
4FGL J0406.0$-$5407 & $<1.1$ & ... & ... & $0.03^{+1.07}_{-0.03}$ & 16.4 & 12.7 & 1.65 & $0.46^{+0.06}_{-0.46}$ & 25.3 & 56.4 & Spi4\_template\_norm.sed\\
4FGL J0439.8$-$1859 & $<1.07$ & ... & ... & $0.85^{+0.22}_{-0.85}$ & 12.3 & 32.2 & 1.1 & $0.00^{+0.03}_{-0.00}$ & 14.1 & 97.8 & I22491\_60\_TQSO1\_40.sed\\
4FGL J0500.6$-$4911 & $...$ & ... & ... & $0.93^{+0.26}_{-0.93}$ & 32.2 & 32.2 & 1.3 & $0.08^{+0.02}_{-0.02}$ & 34.8 & 98.1 & I22491\_80\_TQSO1\_20.sed\\
4FGL J0551.0$-$1622 & $...$ & ... & ... & $...$ & ... & ... & ... & $...$ & ... & ... & ...\\
4FGL J0558.8$-$7459 & $<1.85$ & ... & ... & $1.58^{+0.27}_{-0.14}$ & 2.7 & 75.2 & 1.65 & $1.83^{+0.11}_{-1.83}$ & 6.5 & 73.9 & pl\_QSOH\_template\_norm.sed\\
4FGL J0700.5$-$6610 & $<1.08$ & ... & ... & $0.81^{+0.27}_{-0.81}$ & 5.4 & 31.9 & 1.4 & $0.00^{+0.01}_{-0.00}$ & 7.3 & 99.8 & I22491\_80\_TQSO1\_20.sed\\
4FGL J0746.6$-$4754 & $<1.21$ & ... & ... & $1.00^{+0.21}_{-1.00}$ & 1.4 & 34.9 & 0.95 & $0.01^{+0.03}_{-0.01}$ & 6.1 & 87.1 & I22491\_40\_TQSO1\_60.sed\\
4FGL J0816.9$+$2050 & $...$ & 0.06 & ... & $...$ & ... & ... & ... & $...$ & ... & ... & ...\\
3FGL J0851.8$+$5531 & $<1.21$ & ... & 1.45 & $1.03^{+0.18}_{-1.03}$ & 4.5 & 32.2 & 1.25 & $0.03^{+0.06}_{-0.03}$ & 12.3 & 92.2 & I22491\_70\_TQSO1\_30.sed\\
4FGL J0854.3$+$4408 & $<1.14$ & 0.38 & $2.05^{+0.32}_{-0.32}$ & $0.96^{+0.18}_{-0.96}$ & 0.4 & 32.7 & 0.8 & $0.01^{+0.02}_{-0.01}$ & 4.2 & 71.7 & pl\_I22491\_30\_TQSO1\_70.sed\\
4FGL J0859.4$+$6218 & $<1.1$ & ... & 0.92 & $0.04^{+1.06}_{-0.04}$ & 0.6 & 12.8 & 1.2 & $0.00^{+0.04}_{-0.00}$ & 1.6 & 73.6 & I22491\_60\_TQSO1\_40.sed\\
4FGL J0910.6$+$3329 & $<1.17$ & 0.35 & ... & $1.03^{+0.14}_{-1.03}$ & 1.8 & 32.8 & 0.9 & $0.00^{+0.02}_{-0.00}$ & 6.1 & 90.7 & I22491\_40\_TQSO1\_60.sed\\
4FGL J0910.6$+$3329 & $<4.0$ & ... & ... & $2.73^{+1.27}_{-0.27}$ & 0.0 & 44.7 & 2.0 & $0.09^{+3.91}_{-0.09}$ & 0.0 & 9.4 & S0\_70\_QSO2\_30.sed\\
4FGL J1019.7$+$6321 & $<1.15$ & ... & ... & $0.85^{+0.30}_{-0.85}$ & 13.3 & 30.4 & 1.25 & $0.00^{+0.02}_{-0.00}$ & 11.3 & 87.2 & I22491\_70\_TQSO1\_30.sed\\
4FGL J1059.2$-$1134 & $...$ & ... & 1.34 & $0.25^{+0.59}_{-0.25}$ & 32.1 & 29.7 & 1.65 & $0.00^{+0.03}_{-0.00}$ & 19.9 & 71.5 & I22491\_90\_TQSO1\_10.sed\\
4FGL J1248.3$+$5820 & $<1.06$ & ... & ... & $0.58^{+0.48}_{-0.58}$ & 0.3 & 29.0 & 1.2 & $0.00^{+0.02}_{-0.00}$ & 3.9 & 94.1 & I22491\_60\_TQSO1\_40.sed\\
4FGL J1259.8$-$3749 & $...$ & ... & ... & $1.14^{+0.00}_{-0.08}$ & 90.5 & 39.9 & 3.0 & $3.96^{+0.04}_{-0.00}$ & 251.6 & 100.0 & pl\_QSO\_DR2\_029\_t0.spec\\
4FGL J1304.9$-$2107 & $<1.06$ & ... & ... & $0.73^{+0.33}_{-0.73}$ & 1.1 & 27.7 & 1.5 & $0.13^{+0.58}_{-0.13}$ & 0.1 & 33.5 & CB1\_0\_LOIII4.sed\\
4FGL J1307.6$-$4259 & $<1.34$ & ... & ... & $1.07^{+0.27}_{-0.32}$ & 6.1 & 37.9 & 1.1 & $0.00^{+0.01}_{-0.00}$ & 12.7 & 45.5 & I22491\_60\_TQSO1\_40.sed\\
4FGL J1427.6$-$3305 & $...$ & ... & ... & $1.53^{+0.12}_{-0.15}$ & 72.1 & 78.0 & 2.0 & $0.46^{+0.04}_{-0.04}$ & 10.5 & 98.9 & Sey2\_template\_norm.sed\\
4FGL J1516.8$+$3651 & $<2.15$ & ... & 4.34 & $1.87^{+0.28}_{-0.27}$ & 2.4 & 82.6 & 1.9 & $0.01^{+0.01}_{-0.01}$ & 4.6 & 83.3 & S0\_80\_QSO2\_20.sed\\
4FGL J1612.4$-$3100 & $<1.39$ & ... & ... & $0.98^{+0.41}_{-0.98}$ & 8.0 & 26.6 & 1.5 & $0.16^{+0.06}_{-0.06}$ & 7.1 & 85.4 & Spi4\_template\_norm.sed\\
4FGL J1624.6$+$5651 & $<1.16$ & ... & ... & $0.96^{+0.20}_{-0.96}$ & 13.3 & 30.9 & 1.3 & $0.00^{+0.01}_{-0.00}$ & 15.7 & 90.6 & I22491\_70\_TQSO1\_30.sed\\
4FGL J1707.1$-$1931 & $...$ & ... & ... & $...$ & ... & ... & ... & $4.00^{+0.00}_{-0.18}$ & 0.1 & 62.1 & pl\_QSOH\_template\_norm.sed\\
4FGL J1754.2$+$3212 & $<1.1$ & ... & ... & $0.82^{+0.28}_{-0.82}$ & 1.4 & 31.7 & 1.05 & $0.00^{+0.03}_{-0.00}$ & 4.3 & 89.7 & I22491\_50\_TQSO1\_50.sed\\
3FGL J1829.4$+$5402 & $<1.3$ & ... & ... & $1.14^{+0.16}_{-0.30}$ & 1.0 & 39.6 & 1.0 & $0.00^{+0.03}_{-0.00}$ & 6.8 & 74.6 & I22491\_50\_TQSO1\_50.sed\\
4FGL J1836.4$+$3137 & $<1.15$ & ... & ... & $0.87^{+0.28}_{-0.87}$ & 10.2 & 30.6 & 1.8 & $0.00^{+0.03}_{-0.00}$ & 11.1 & 57.4 & Sey18\_template\_norm.sed\\
4FGL J1838.8$+$4802 & $<1.13$ & 0.30 & ... & $0.92^{+0.21}_{-0.92}$ & 1.1 & 35.2 & 0.85 & $0.01^{+0.02}_{-0.01}$ & 5.3 & 84.5 & pl\_I22491\_30\_TQSO1\_70.sed\\
3FGL J1926.8$+$6154 & $<1.14$ & ... & ... & $0.93^{+0.21}_{-0.93}$ & 1.6 & 32.5 & 0.7 & $0.03^{+0.21}_{-0.02}$ & 3.7 & 27.8 & pl\_I22491\_10\_TQSO1\_90.sed\\
4FGL J1946.0$-$3112 & $<1.48$ & ... & ... & $1.38^{+0.10}_{-0.11}$ & 15.3 & 70.4 & 1.2 & $0.10^{+0.05}_{-0.05}$ & 15.5 & 95.4 & Spi4\_template\_norm.sed\\
4FGL J1959.7$-$4725 & $<1.06$ & ... & ... & $0.26^{+0.80}_{-0.26}$ & 4.6 & 23.1 & 1.05 & $0.00^{+0.01}_{-0.00}$ & 5.9 & 89.7 & I22491\_50\_TQSO1\_50.sed\\
4FGL J2030.9$+$1935 & $<0.96$ & ... & ... & $0.03^{+0.93}_{-0.03}$ & 8.0 & 14.6 & 1.05 & $0.04^{+0.09}_{-0.04}$ & 6.7 & 82.9 & I22491\_50\_TQSO1\_50.sed\\
4FGL J2126.1$-$3922 & $<2.1$ & ... & ... & $1.77^{+0.33}_{-0.43}$ & 0.3 & 63.6 & 1.45 & $1.83^{+0.02}_{-0.53}$ & 2.4 & 45.5 & pl\_QSOH\_template\_norm.sed\\
4FGL J2149.6$+$0323 & $<1.34$ & ... & $1.28$ & $1.15^{+0.19}_{-0.57}$ & 0.5 & 37.5 & 1.4 & $0.00^{+0.02}_{-0.00}$ & 9.9 & 93.7 & I22491\_90\_TQSO1\_10.sed\\
4FGL J2243.7$-$1231 & $<1.05$ & 0.23 & ... & $0.03^{+1.02}_{-0.03}$ & 2.5 & 13.2 & 0.75 & $0.53^{+0.09}_{-0.53}$ & 1.4 & 58.3 & pl\_QSOH\_template\_norm.sed\\
4FGL J2243.9$+$2021 & $<1.18$ & ... & ... & $1.00^{+0.18}_{-1.00}$ & 0.3 & 36.9 & 0.95 & $0.01^{+0.02}_{-0.01}$ & 5.9 & 94.4 & I22491\_40\_TQSO1\_60.sed\\
4FGL J2325.6$+$1644 & $<1.04$ & ... & ... & $0.03^{+1.01}_{-0.03}$ & 4.7 & 13.7 & 0.9 & $0.18^{+0.06}_{-0.18}$ & 3.0 & 56.8 & I22491\_40\_TQSO1\_60.sed\\
4FGL J2352.0$+$1750 & $<1.53$ & ... & ... & $1.34^{+0.19}_{-0.19}$ & 9.3 & 59.3 & 0.95 & $1.29^{+0.06}_{-0.08}$ & 7.2 & 99.8 & pl\_QSOH\_template\_norm.sed\\
3FHL J2358.4$-$1808 & $<1.11$ & 0.06 & ... & $0.85^{+0.26}_{-0.85}$ & 6.3 & 31.6 & 1.25 & $0.00^{+0.01}_{-0.00}$ & 6.4 & 98.0 & I22491\_70\_TQSO1\_30.sed\\
3FGL J0707.2$+$6101 & $<1.25$ & ... & ... & $1.06^{+0.19}_{-1.06}$ & 11.0 & 36.1 & 2.0 & $0.00^{+0.02}_{-0.00}$ & 19.2 & 95.8 & Mrk231\_template\_norm.sed\\
4FGL J0800.9$+$4401 & $<2.13$ & 1.07 & 1.38 & $1.90^{+0.23}_{-0.21}$ & 4.0 & 83.9 & 1.3 & $0.02^{+0.02}_{-0.02}$ & 8.7 & 21.5 & S0\_90\_QSO2\_10.sed\\
3FGL J0103.4$+$5336 & $<1.05$ & ... & ... & $0.04^{+1.01}_{-0.04}$ & 14.0 & 12.8 & 2.0 & $0.07^{+0.04}_{-0.04}$ & 3.7 & 99.9 & Mrk231\_template\_norm.sed\\
3FGL J1138.2$+$4905 & $...$ & ... & 1.30 & $...$ & ... & ... & ... & $2.64^{+0.24}_{-0.18}$ & 8.5 & 81.5 & Mrk231\_template\_norm.sed\\
3FGL J1323.9$+$1405 & $<1.38$ & 0.47 & 0.77 & $1.24^{+0.14}_{-0.13}$ & 3.1 & 68.9 & 0.65 & $1.21^{+0.07}_{-0.09}$ & 3.6 & 92.2 & pl\_I22491\_30\_TQSO1\_70.sed\\
3FGL J1354.5$+$3705 & $<1.0$ & 0.38 & ... & $0.24^{+0.76}_{-0.24}$ & 4.5 & 24.4 & 1.65 & $0.00^{+0.01}_{-0.00}$ & 18.9 & 76.4 & I22491\_90\_TQSO1\_10.sed\\
3FGL J1500.6$+$4750 & $<2.06$ & 1.06 & $0.77^{+1.25}_{-1.25}$ & $1.84^{+0.22}_{-0.31}$ & 10.5 & 65.9 & 0.25 & $1.40^{+0.11}_{-0.40}$ & 11.1 & 56.8 & pl\_QSO\_DR2\_029\_t0.spec\\
3FGL J1649.4$+$5238 & $<1.02$ & ... & ... & $0.12^{+0.90}_{-0.12}$ & 2.8 & 21.5 & 1.6 & $0.00^{+0.03}_{-0.00}$ & 11.4 & 92.2 & I22491\_90\_TQSO1\_10.sed\\
3FGL J1702.6$+$3116 & $<1.45$ & 0.47 & $1.17^{+0.36}_{-0.36}$ & $1.27^{+0.18}_{-0.22}$ & 15.3 & 52.4 & 0.9 & $1.23^{+0.07}_{-0.06}$ & 13.9 & 98.9 & pl\_QSOH\_template\_norm.sed\\
\enddata
    \end{deluxetable*}

\vspace{-0.2cm}
\begin{footnotesize}
$^a$ Best-fit or upper limit of the photometric redshift, either a value or an upper limit.

$^b$ The spectroscopic redshifts are extracted from \cite{Ajello_2020}, \cite{sdss_dr7} and \cite{marchesini2019optical} unless specifically denoted.

$^c$ Photometric redshifts with 1$\sigma$ confidence level.

$^d$ $\chi^2$ value for ten degrees of freedom.

$^e$  Redshift probability density at $z_{\text {phot}} \pm 0.1\left(1+z_{\text {phot }}\right)$.

$^f$ Spectral slope for the power-law model of the form $F_\lambda \propto \lambda^{-\beta}$.

$^h$ The spectroscopic redshift is from \cite{truebenbachVLBAExtragalacticProper2017}.

\end{footnotesize}

\section{Results} \label{sec:redults}

\begin{figure}[h]
    \centering
    \includegraphics[width=\textwidth]{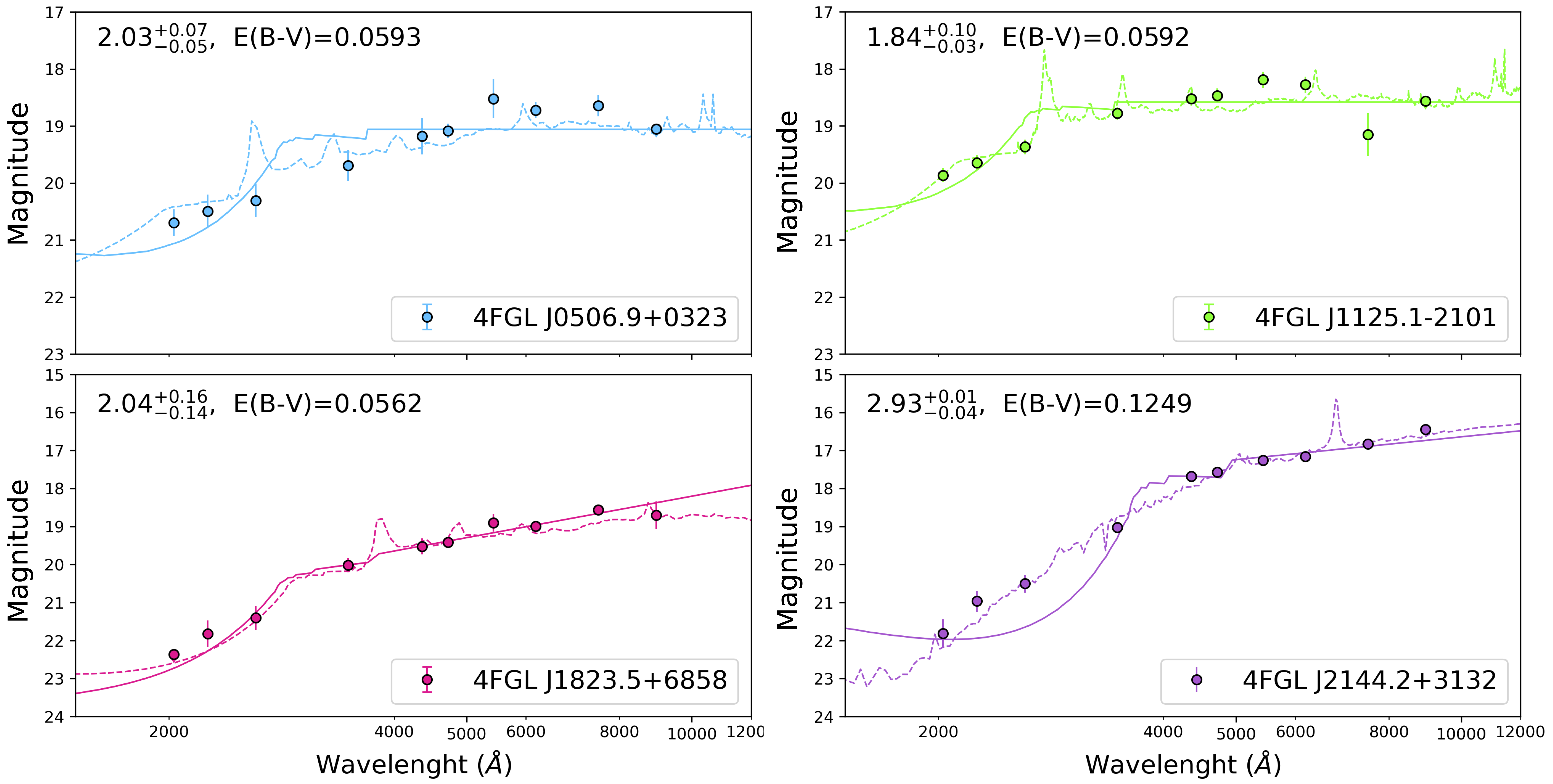}
    %\captionsetup{width=0.8\textwidth}
    \caption{The SEDs of the four high-redshift BL Lacs. The $x$ axis is the wavelength while the y axis is the magnitudes. The solid circles with black edges are photometric data from Table \ref{tab:result}. The magnitudes are ordered in the following the central wavelength of the filters: $uw2,\ um2,\ uw1,\ uuu,\ ubb,\ g',\ uvv,\ r',\ i',\ z'$. The solid line represents power-law model while dashed line represents a galaxy model.}
    \label{fig:high-$z$_sed}
\end{figure}

Table \ref{tab:result} reports the photometric redshifts or upper limits, fitting statistics, and the corresponding SED models for the sample. Four high-$z$ BL Lacs are reported, and detailed information about the models and magnitudes is presented in Figure \ref{fig:high-$z$_sed}. 
Two sources, 4FGL J0506.9$+$0323 and 4FGL J1823.5$+$6858, are best fit by power-law models, yielding photometric redshifts at $2.03^{+0.07}_{-0.05}$ and $2.04^{+0.16}_{-0.14}$, respectively. For the other two sources, 4FGL J1125.1$-$2101 and 4FGL J2144.2$+$3132, the fit slightly prefers Type I AGN (pl\_QSO\_DR2\_029\_t0) and a mix of a lenticular galaxy and partially obscured Type I AGN (S0\_70\_QSO2\_30).
%suggesting a possible contribution from the host galaxy that results in broad emission features in the optical band. 
However, their optical spectra from ZBLLAC \citep{landoniZBLLACSpectroscopicDatabase2020} are featureless and well described by power-law continuum, consistent with their BL Lac classification. Therefore, the preferred galaxy/AGN fits might be caused by template degeneracy, and we report the redshifts derived from the power-law fits at $1.84^{+0.10}_{-0.03}$ and $2.93^{+0.01}_{-0.04}$, which we consider more physically appropriate for these sources. The best-fit optical power-law indices for the four high-$z$ sources are consistent with the optical spectral indices of 475 BL Lacs reported by \cite{shawSPECTROSCOPYLARGESTEVER2013a}, which span the range -0.291 to 2.250 after excluding outliers using the Median Absolute Deviation (MAD) method.

We conducted a comprehensive literature search of spectroscopic redshifts from several major databases, including NASA/IPAC Extragalactic Database (NED) \citep{chenBestPracticesData2022}, SDSS Data Release (DR) 10 \citep{ahnTenthDataRelease2014}, DESI DR10 \citep{deyOverviewDESILegacy2019}, ZBLLAC \citep{landoniZBLLACSpectroscopicDatabase2020a}, and Roma-BZCAT \citep{massaroRomaBZCATMultifrequencyCatalogue2009, massaro5thEditionRomaBZCAT2015} for the four high-$z$ BL Lacs we found. The search shows that there is only one
measured the spectroscopic redshift of 4FGL J1823.5$+$6858 at $2.140^{+0.001}_{-0.001}$ \citep{VLBA_redshifts} , which is compatible with our result at $2.04^{+0.16}_{-0.14}$. 
\cite{dominguez2024constraints} provides the redshift upper limits of blazars from the EBL attenuation. The photometric redshift of 4FGL J1125.1$-$2101, $z=1.84^{+0.10}_{-0.03}$, is compatible with the upper limit $z<1.76^{+0.90}_{-0.62}$ from \cite{dominguez2024constraints}. Another source, 4FGL J2144.2$+$3132, has a photometric redshift at $z=2.93^{+0.01}_{-0.04}$, while the upper limit is only compatible with it at the 2$\sigma$ level \cite{dominguez2024constraints}.
%is $z<0.79^{+0.78}_{-0.74}$ from \cite{dominguez2024constraints}.

The magnitudes of three high-$z$ sources with $z<2.1$ were converted to the $R$ band using the transformation equations in \cite{jesterSloanDigitalSky2005}. The absolute $R$-band magnitude was then computed as
$$
M = m-\mu-K(z),
$$
where $m$ is the apparent magnitude, $\mu$ is the distance modulus, and $K(z) = 2.5(\beta -1)\log_{10}(1+z)$ is the $K$-correction for a power-law spectrum $F_\nu \propto \nu^{-\beta}$. 
For the three sources, we found $M_R = -26.15,\ -26.11,\ \text{and } -27.73$ for 4FGL J0506.9+0323, 4FGL J1125.1$−$2101, and 4FGL J1823.5+6858, respectively. These values are significantly brighter than the typical BL Lac host-galaxy luminosity ($M_R = -22.8,\ \sigma=0.5$; \citealt{sbarufattiImagingRedshiftsBL2005}). 
Moreover, the derived $M_R$ values are consistent with those of BL Lac nuclei, which range from $M_R \sim -28.7$ to $-18.5$ reported in \cite{sbarufattiImagingRedshiftsBL2005}. This indicates that the optical emission of these high-$z$ sources is dominated by the nucleus rather than by the host galaxy.

Metal-rich structures along the line of sight, such as foreground galaxies and intergalactic gas clouds, can imprint characteristic absorption features onto the spectra of background BL Lacs \citep{bergeronSampleGalaxiesGiving1991, steidelQuasarAbsorbingGalaxies1997}. Among these features, the \ion{Mg}{2} $\lambda\lambda$ 2796, 2803 $\AA$ doublet is the most commonly studied metallic narrow absorption line, and such systems are generally referred as \ion{Mg}{2} absorbers. To estimate the probability of not detecting any intervening \ion{Mg}{2} absorbers in our sources, we utilize the two available optical spectra with reported signal-to-noise ratios (S/N): 4FGL J1125.1$-$2101 \citep{herazoOpticalSpectroscopicObservations2017, landoniZBLLACSpectroscopicDatabase2020a} and 4FGL J2144.2$+$3132 \citep{marchesiniOpticalSpectroscopicObservations2019, landoniZBLLACSpectroscopicDatabase2020a}. Depending on the S/N of the spectra, we adopt the incidence rate ($\mathrm{d}N/\mathrm{d}z$) for strong (rest-frame equivalent width, $W_0\geq1\AA$) and weak ($W_0\geq0.3\AA$) absorption systems respectively: 
\cite{mishraIncidenceMgIi2018} reports $\mathrm{d}N/\mathrm{d}z=0.33_{-0.14}^{+0.22}$ for weak system and $\mathrm{d}N/\mathrm{d}z=0.23_{-0.09}^{+0.13}$ for strong systems in Sample III of BL Lacs.
4FGL J1125.1$-$2101 has an optical spectrum covering the wavelength range 4000-7000 $\AA$, corresponding to a redshift path of $\Delta z = 1.07$. Given the high spectrum S/N of 103, we apply the incidence rate for weak systems. This yields an absorber incidence expectation of $\mu=0.34^{+0.24}_{-0.15}$, computed as $\mu=(d N / d z) \times \Delta z$. Assuming the incidences are independent Poisson events, the probability of non-detection of \ion{Mg}{2} absorbers is $P(k=0,\ \mu) =70.3^{+11.4}_{-14.7}\%$. A similar calculation was performed for 4FGL J2144.2$+$3132, whose optical spectrum spans 3800-8000 $\AA$ and has an S/N of 23, which leads to the incidence rate for strong systems. The calculations yields a probability of $70.8^{+10.2}_{-12.6}\%$. Therefore, the joint probability of not detecting \ion{Mg}{2} absorbers from both spectra is $49.8^{+16.4}_{-17.4}\%$. This suggests that the absence of \ion{Mg}{2} absorbers in the two spectra is not unexpected, given the limited redshift path of the optical spectra and the statistical incidence of such systems along BL Lac sightlines.
%This suggests that the absence of detectable intervening \ion{Mg}{2} absorption in the two spectra is not unexpected given the limited redshift path and the statistical incidence of such systems along BL Lac sightlines.

\section{Discussion} \label{sec: discussions}
\subsection{Cosmic Gamma-ray Horizon}

The EBL interacts with the very-high-energy (VHE) photon ($>$100 GeV) from the blazars. The EBL photons from ultraviolet to infrared band annihilate with the VHE photons, impeding the travel of VHE photon in the universe. 
The cosmic gamma-ray horizon (CGRH, \cite{dominguezDETECTIONCOSMICUpgamma2013} is the distance at which, for a given energy, the universe is opaque to the propagation of gamma rays of that energy \citep{Finke_model, Dominguez_CGRH}.
%Therefore, there is a redshift-dependent opacity for VHE photons to each us \citep{Finke_model, Dominguez_CGRH}, which is called the cosmic gamma-ray horizon (CGRH, \cite{dominguezDETECTIONCOSMICUpgamma2013}).

\begin{figure}[h]
\centering
\includegraphics[width=0.7\textwidth]{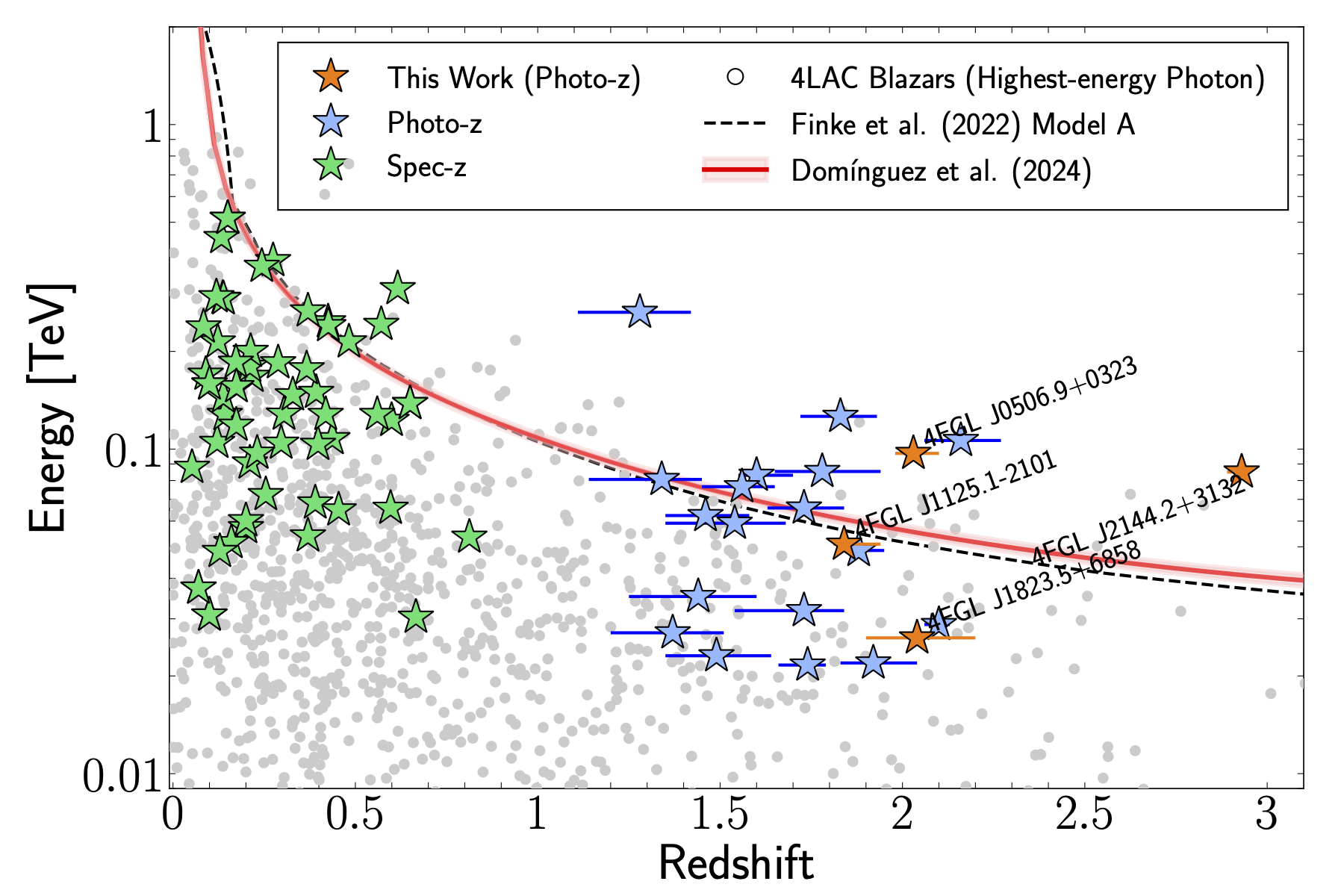}
%\captionsetup{width=0.8\textwidth}
\caption{The cosmic gamma-ray horizon plot. The grey circles are from the 4LAC blazars. 
%The line with errors represent the horizon ($\tau$=1) according to \cite{dominguez2024new_ebl_model}.
The blue and orange stars are the high-$z$ BL Lacs found by the Photo-$z$ campaign \citep{Rau2012,kaur2017,kaur2018,Rajagopal_2020, sheng2024revealing} including this work. Note that only high-$z$ BL Lac reported with highest energy photon (HEP) are plotted here.
The green stars are BL Lacs with spectroscopic redshift determined by \cite{marchesi2018identifying,desai2019identifying, paiano2020optical, rajagopal2021identifying, goldoni2021optical, garcia2023optical, kasai2023optical}.
%, which are blazars of uncertain type (BCU).} 
}
\label{fig:cgrh}
\end{figure}

Figure \ref{fig:cgrh} shows the EBL model \citep{saldana2021observational, finke2022modeling, dominguez2024new_ebl_model} plotted with the highest energy photons from 4LAC blazars, blazars with spectroscopic redshifts, and the high-$z$ BL Lacs found by the photo-$z$ campaign. 
Based on the photometric redshifts measured in this work and the highest photon energies reported in the 4LAC catalog, we obtained optical depth of $0.74,\ 0.04,\ 2.40,\ \text{and}\ 2.93$ for 4FGL J1125.1$−$2101, 4FGL J1823.5+6858, 4FGL J0506.9+0323, and 4FGL J2144.2+3132, respectively. Two sources exhibit negligible-to-moderate absorption ($\tau<1$), making their detections fully consistent with the EBL transparency. The other two have $\tau>2$, which implies significant attenuation, making observed fluxes reduced to $\sim9\%$ and $\sim5\%$ of their intrinsic values. This may suggest a more transparent EBL model at high redshift, although a larger statistical sample of high-$z$ sources will be needed to draw firmer conclusions about EBL transparency.
The high-$z$ BL Lacs expand the population on the high-redshift side of the CGRH plot, helping to constrain the EBL model in a regime where data points are scarce. Future discoveries of additional high-$z$ BL Lacs will be essential for more robust tests of EBL transparency at high redshift.
%The high-$z$ BL Lacs have increased the population of the sources on the high-redshift side of the CGRH plot, thus constraining the EBL model where data points are scarce. 

\begin{figure}[h]
    \centering
    \includegraphics[width=0.5\textwidth]{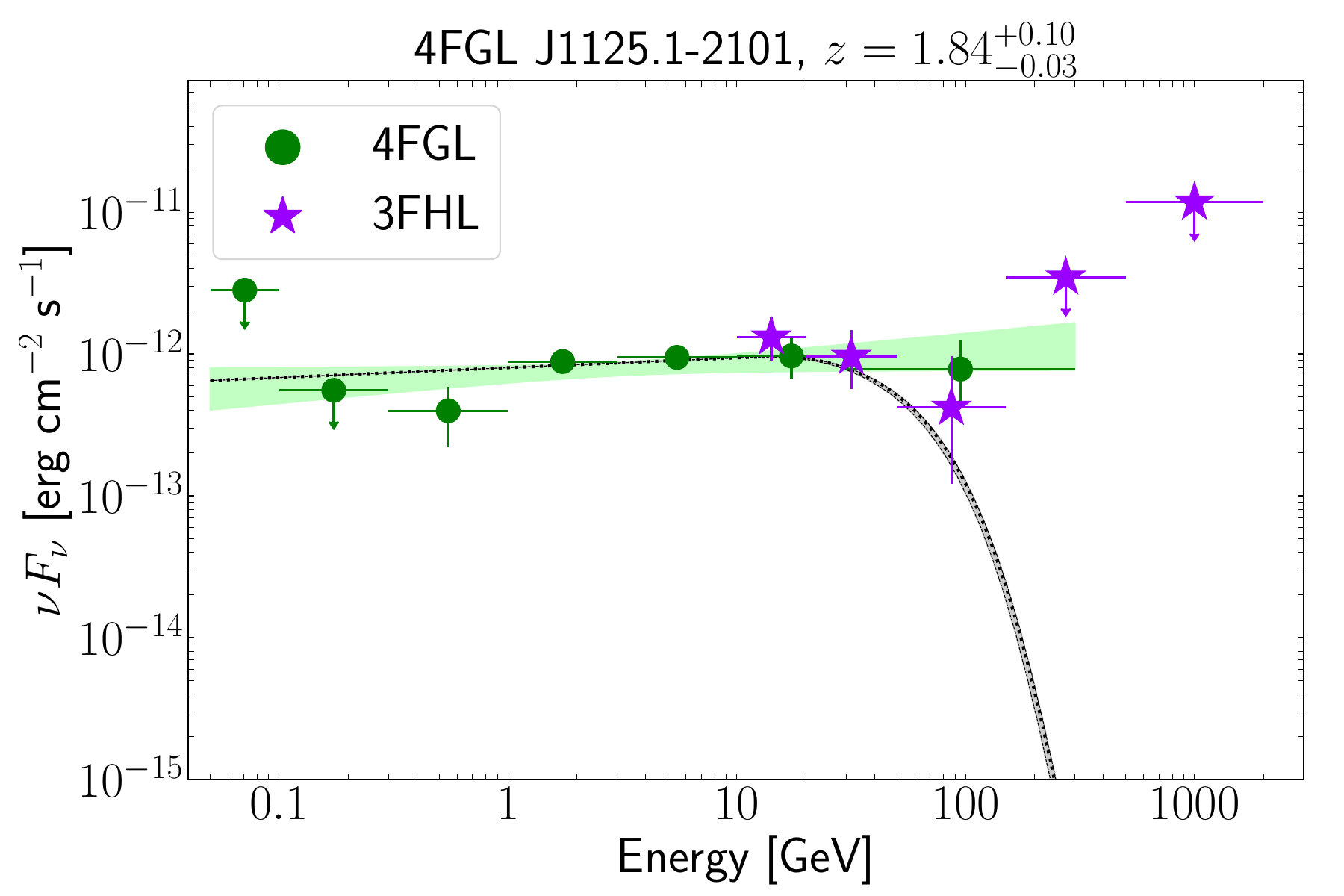}\hfill
    \includegraphics[width=0.5\textwidth]{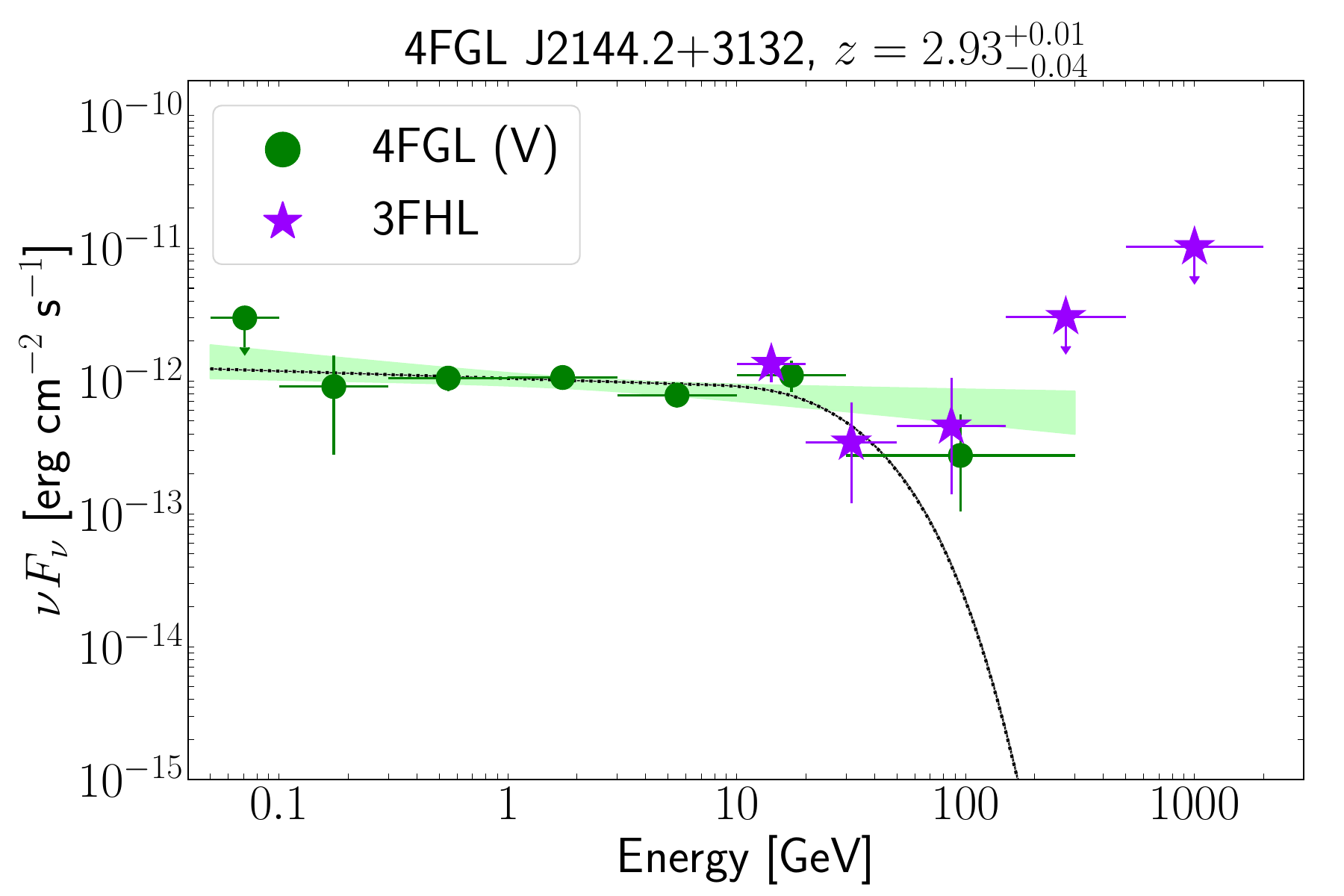}\hfill
    \includegraphics[width=0.5\textwidth]{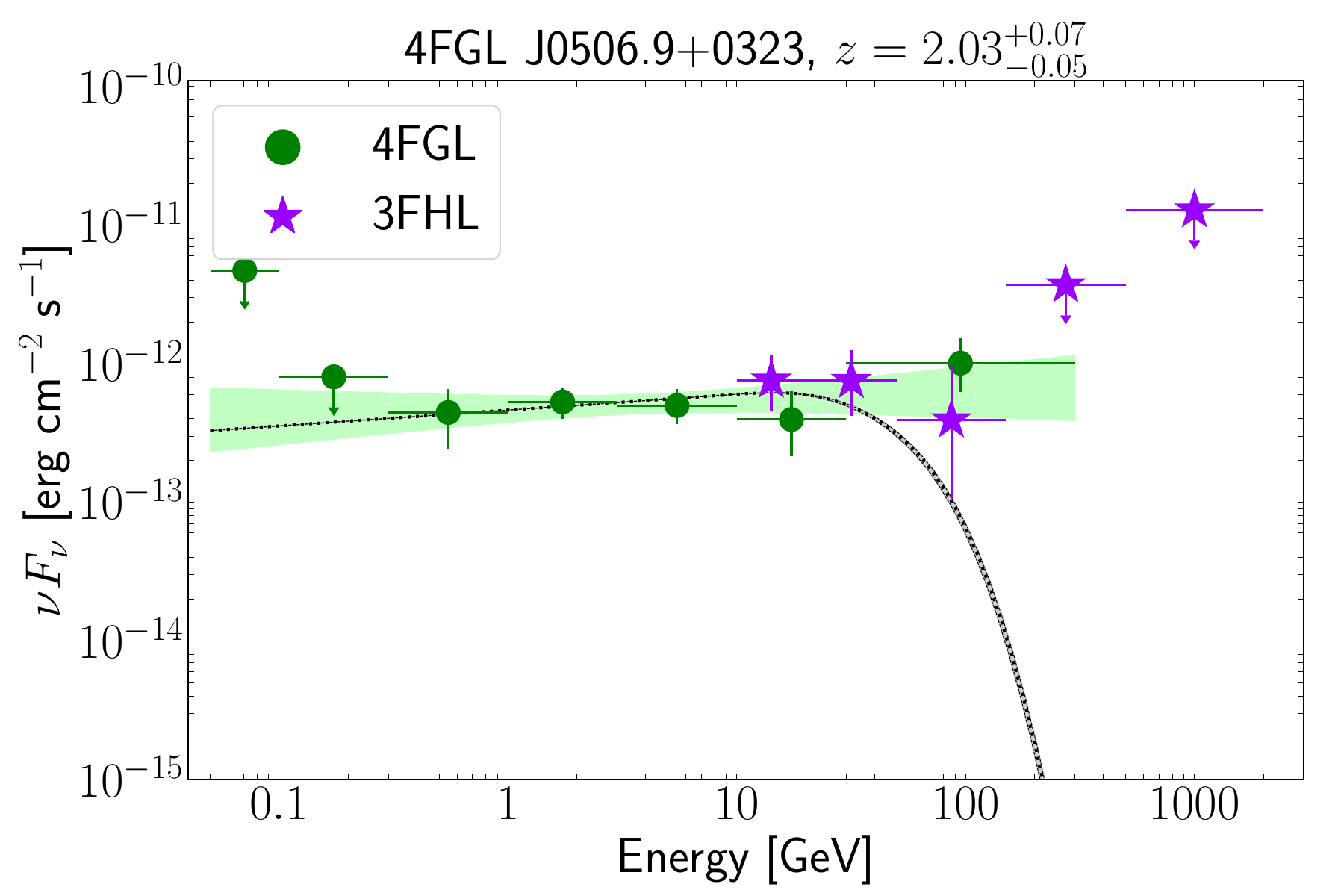}\hfill
    \includegraphics[width=0.5\textwidth]{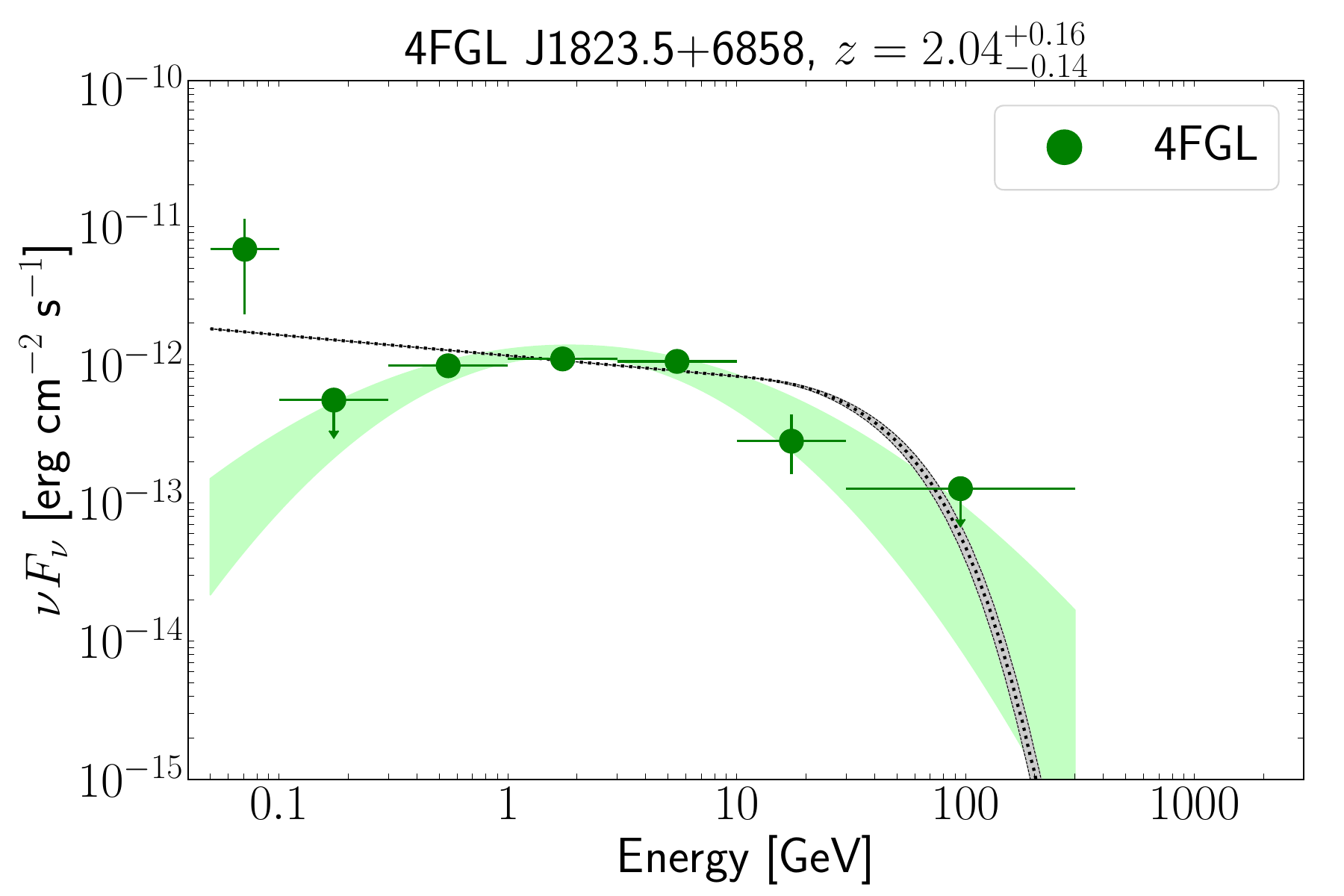}\hfill
    % \captionsetup{width=0.9\textwidth}
    \caption{The powerlaw fit to the SED data. The data are obtained from the 4FGL and 3FHL catalogs and plotted using green and purple dots, respectively. The 3FHL data for 4FGL J1823.5$+$6858 is not available. Therefore, only the 4FGL data is shown in the plot. The green bands are the uncertainties from the \emph{Fermi}-LAT data reduction and include statistical uncertainties. The black dotted curve is the best joint power-law fit, which is attenuated by $e^{-\tau(E,z)}$ due to the EBL absorption \citep{saldana2021observational, finke2022modeling, dominguez2024new_ebl_model}, showing the energy at where the cutoff is. The redshift uncertainty has a small effect on the SED fit, indicated by the gray-shaded uncertainty region of the SED curve. }
    \label{fig:sed}
    
\end{figure}

Figure \ref{fig:sed} presents the gamma-ray SEDs for the four high-$z$ sources discovered in this work. We fit the 4FGL and, when available, 3FHL data with a power law attenuated by the EBL absorption factor $e^{-\tau(E,z)}$ \citep{Dominguez_CGRH}. The uncertainty (shaded area) in the fits arises from the photo-$z$ errors, demonstrating that redshift uncertainty has a negligible impact on the source flux. The effect of EBL attenuation is evident in the fitted models.

%The gamma-ray SEDs of the three high-$z$ sources are shown in Figure \ref{fig:sed}. We fit the Fermi-LAT Fourth Source Catalog (4FGL, \citealt{4fgl,4fgl_dr2}) and 3FHL data with a power law  attenuated by EBL absorption. The EBL absorption \citep{Dominguez_CGRH} is applied with a factor $e^{-\tau(E,z)}$ in the fits. The redshift uncertainty determined by the $\chi^2$ fitting is considered in the fits, which is demonstrated by the gray shaded area in the plots. The fits show that redshift uncertainty has a negligible effect on the source flux. Moreover, the effect of redshift uncertainty decreases with higher redshift. In addition, the attenuation caused by the EBL can be seen in the joint fits of the 3FHL and 4FGL data.

%The gamma-ray SEDs of the three high-$z$ sources are shown in Figure \ref{fig:sed}. We fit the Fermi-LAT Fourth Source Catalog (4FGL, \citealt{4fgl,4fgl_dr2}) and 3FHL data with a power law  attenuated by EBL absorption. The EBL absorption \citep{Dominguez_CGRH} is applied with a factor $e^{-\tau(E,z)}$ in the fits. The redshift uncertainty determined by the $\chi^2$ fitting is considered in the fits, which is demonstrated by the gray shaded area in the plots. The fits show that redshift uncertainty has a negligible effect on the source flux. Moreover, the effect of redshift uncertainty decreases with higher redshift. In addition, the attenuation caused by the EBL can be seen in the joint fits of the 3FHL and 4FGL data.

\begin{figure}[!htb]
    \centering
    \subfigure[Synchrotron peak luminosity vs. rest-frame synchrotron peak frequency.]
    { \includegraphics[width=0.48\textwidth]{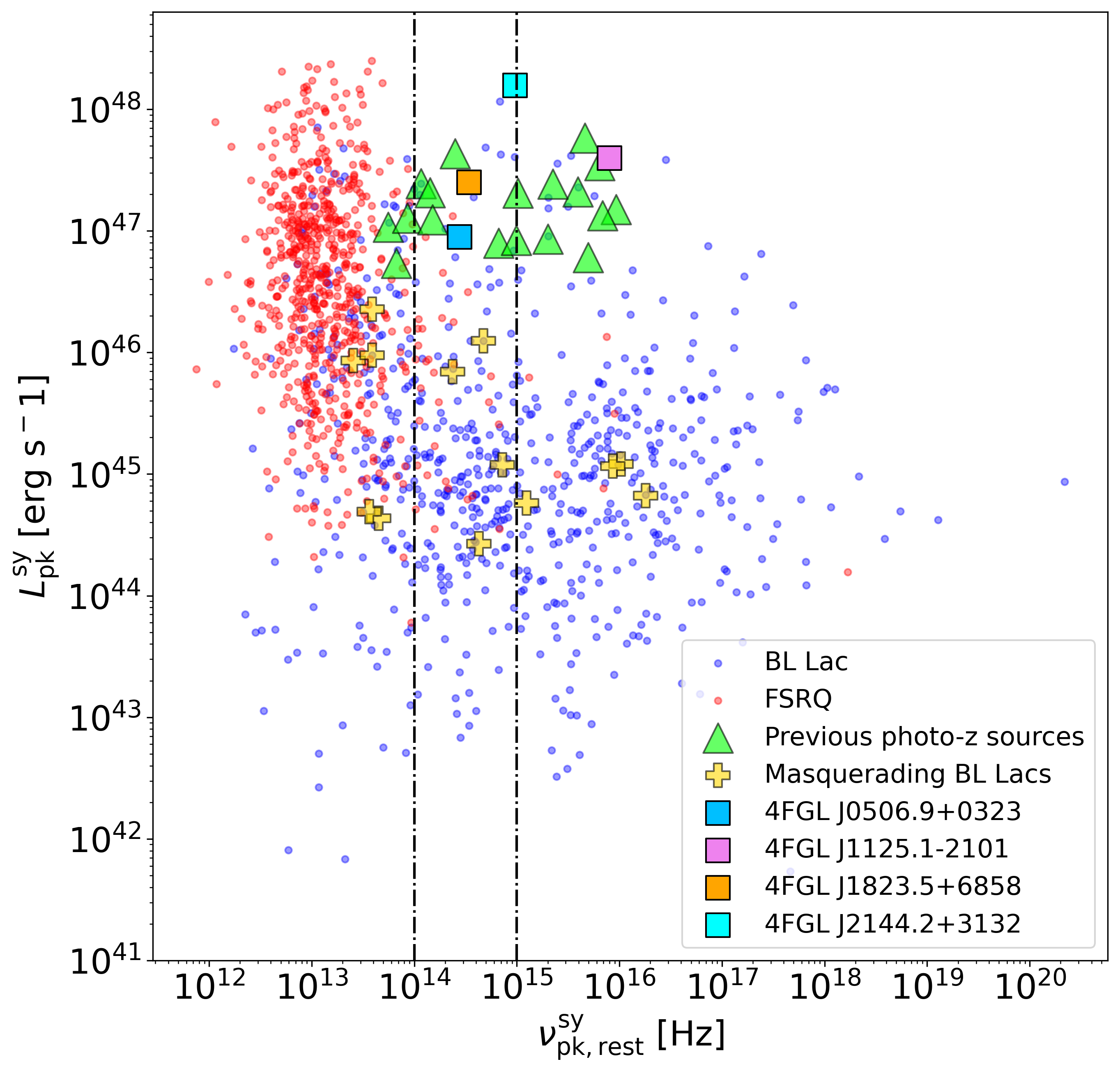}
    \label{fig:sequence_syn_lumi} }\hfill
    \subfigure[Compton dominance vs. synchrotron peak frequency at rest frame.]
    { \includegraphics[width=0.48\textwidth]{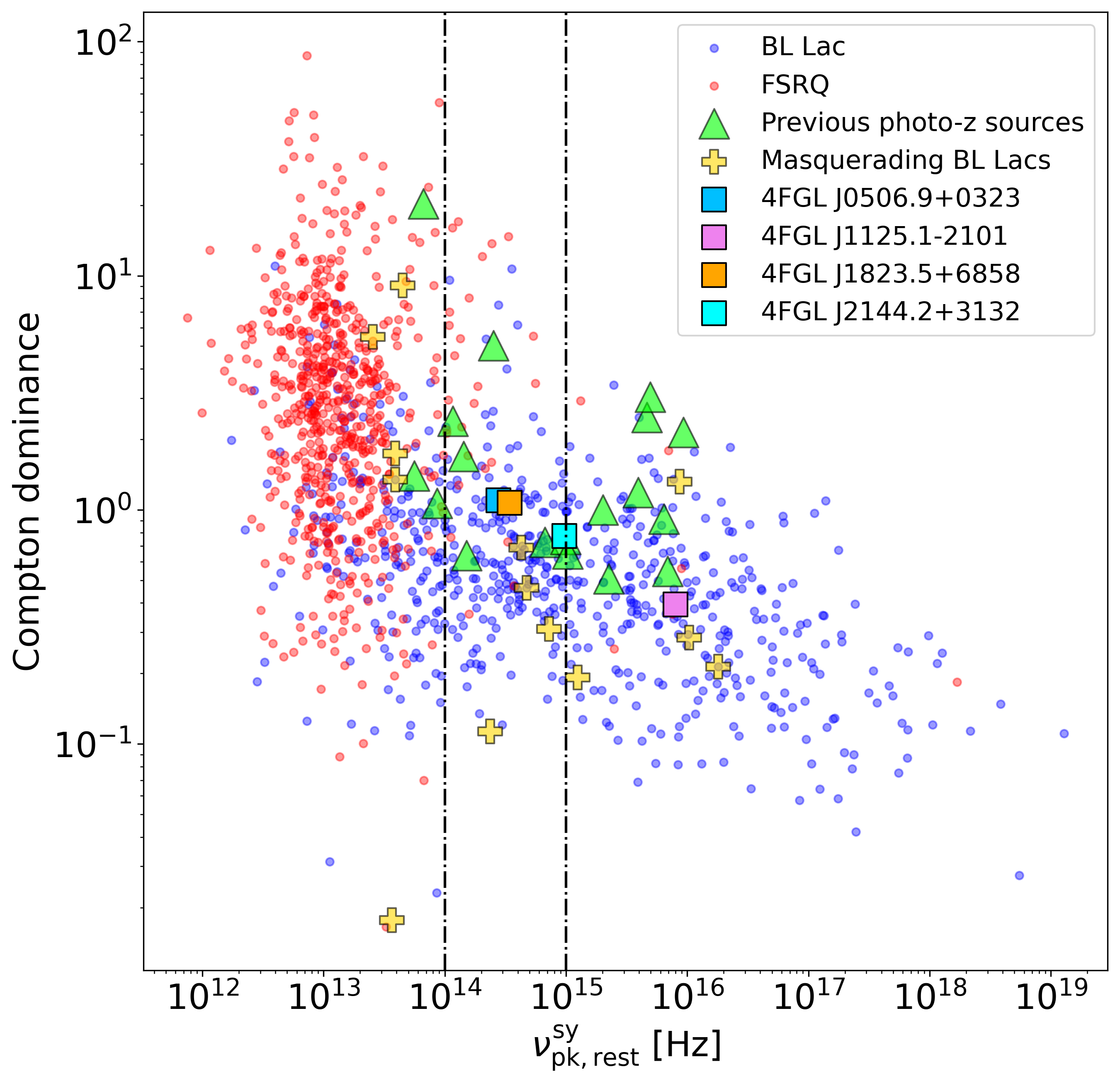}
    \label{fig:sequence_cd} }\\
    \subfigure[Gamma-ray index vs. synchrotron peak frequency at rest frame.]
    { \includegraphics[width=0.48\textwidth]{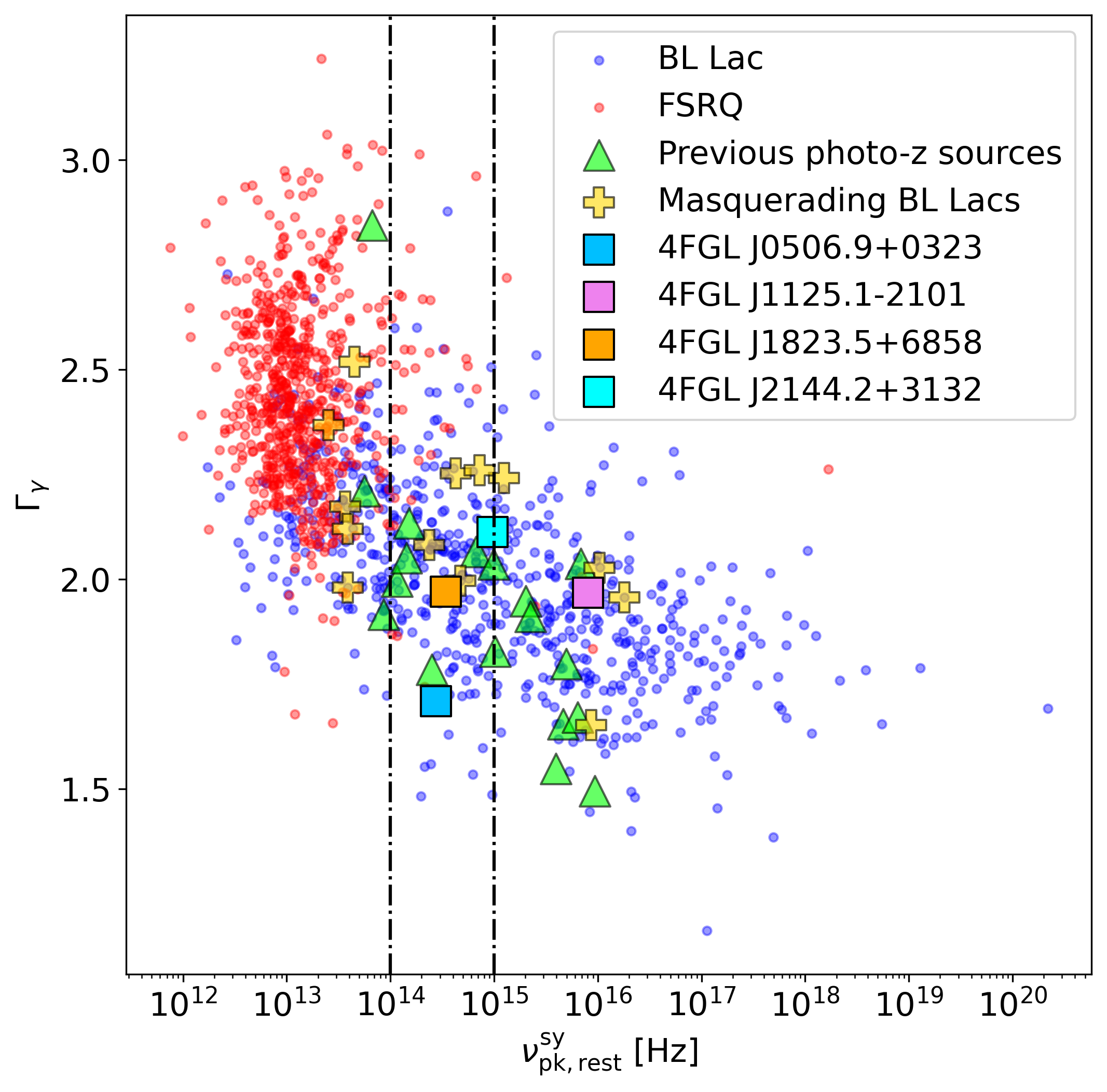}
    \label{fig:sequence_index} }
    \subfigure[Gamma-ray index vs. gamma-ray luminosity (0.1-100 GeV).]
    { \includegraphics[width=0.48\textwidth]{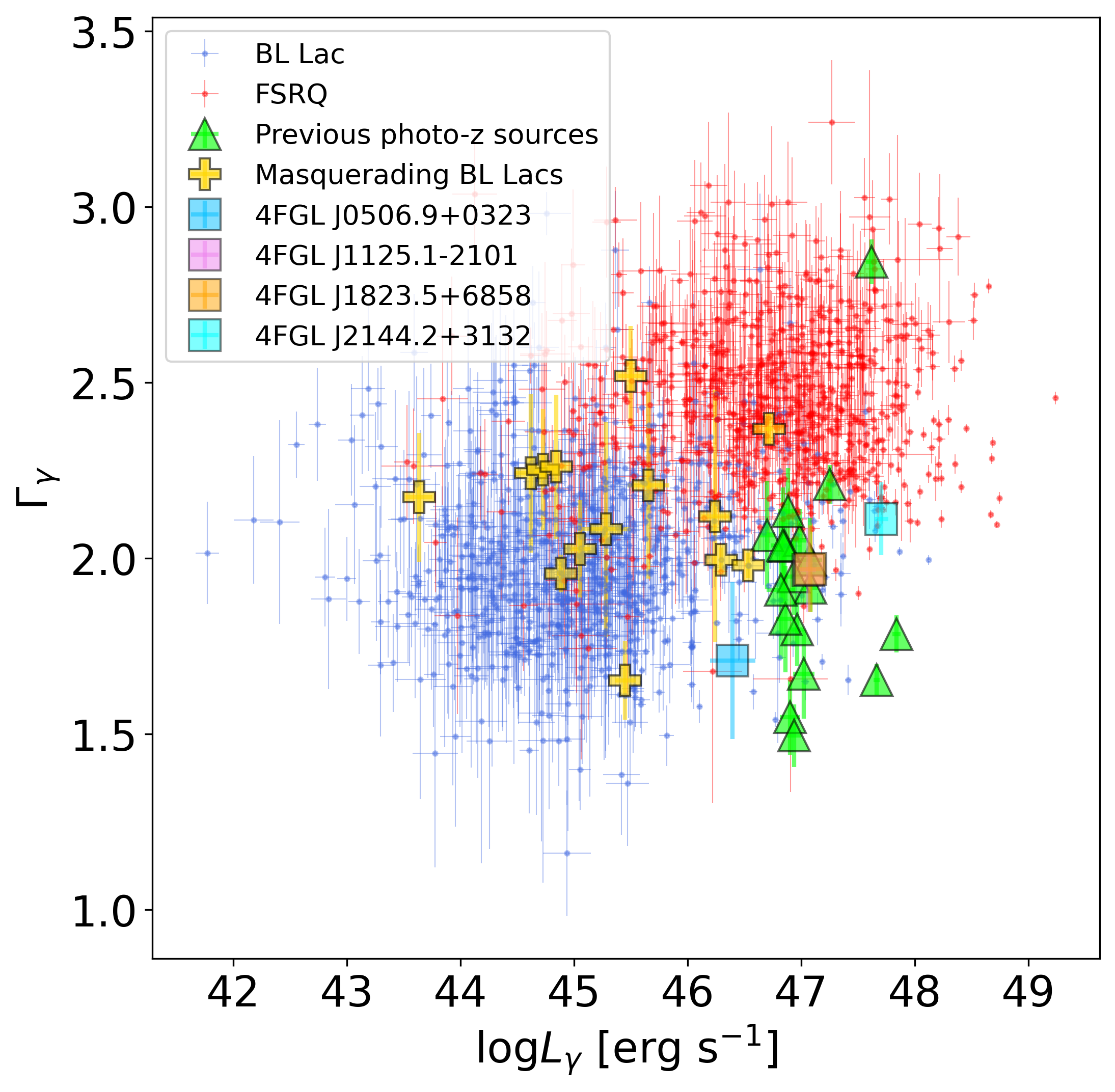}
    \label{fig:blazar_divide} }
\caption{The blazar sequence and Fermi blazar divide plot. The colored dots are calculated from the 4LAC catalog \citep{Ajello_2020}. The triangles are the previous high-$z$ BL Lacs \citep{Rau2012,kaur2017,kaur2018,Rajagopal_2020}, while the squares are the ones found in this work. The vertical dotted lines divides the blazars to LSP, ISP, and HSP groups. The orange crosses are the masquerading BL Lacs from \cite{mbl_lacs, paiano2023spectra, sahakyan2023multimessenger}. The indices are not corrected for the EBL absorption. Note that in (d), 4FGL J1125.1$-$2101 overlaps with 4FGL J1823.5$+$6858.} 
\label{fig:sequence_divide}
\end{figure}

\subsection{Blazar Sequence}
A unified blazar model named blazar sequence is proposed by \cite{fossati1998unifying} to explain the population of blazars. Blazars are separated into two groups, BL Lac and FSRQ, by their spectral properties: the synchrotron peak frequency at rest frame ($\nu^{\mathrm{sy}}_{\mathrm{pk,\ rest}}$) versus synchrotron peak luminosity ($L_{\mathrm{pk}}^{\mathrm{sy}}$), Compton dominance (CD), and gamma-ray index ($\Gamma_{\gamma}$). Compton dominance is the ratio between the inverse-Compton luminosity ($L_{\mathrm{pk}}^{\mathrm{IC}}$) to $L_{\mathrm{pk}}^{\mathrm{sy}}$. The blazar sequence predicts that $\nu^{\mathrm{sy}}_{\mathrm{pk,\ rest}}$ is anti-correlated with $L_{\mathrm{pk}}^{\mathrm{sy}}$, Compton dominance, and $\Gamma_{\gamma}$. BL Lacs dominate areas with higher $\nu^{\mathrm{sy}}_{\mathrm{pk,\ rest}}$, lower $L_{\mathrm{pk}}^{\mathrm{sy}}$, Compton dominance, and gamma-ray index. The FSRQ population shows the opposite relation: they have lower $\nu^{\mathrm{sy}}_{\mathrm{pk,\ rest}}$, higher $L_{\mathrm{pk}}^{\mathrm{sy}}$, Compton dominance, and gamma-ray index. Figure \ref{fig:sequence_syn_lumi}, \ref{fig:sequence_cd}, and \ref{fig:sequence_index} show how the anti-correlation of the spectral properties divides blazars into two population groups. However, the blazar sequence can be caused by some selection effects \citep{giommi2012simplified}. Figure \ref{fig:sequence_syn_lumi} shows a disagreement with the current blazar sequence model. The luminous high-$z$ BL Lacs are found by the photo-$z$ campaign. 
Therefore, our high-$z$ BL Lacs are essential to study the blazar population from an unbiased perspective.

We calculate the spectral properties ($\nu^{\mathrm{sy}}_{\mathrm{pk,\ rest}}$, $L_{\mathrm{pk}}^{\mathrm{sy}}$, CD, $L_{\mathrm{pk}}^{\mathrm{IC}}$) using the 4LAC catalog data. For some high-$z$ sources lacking the synchrotron peak frequency and flux in the 4LAC catalog, we use the {\it SDSS} Sky Explorer to plot the SED and fit the synchrotron and inverse-Compton peak.
%Most of the spectral properties are calculated using the 4LAC catalog. We also utilized two additional methods if the 4LAC catalog does not have the the synchrotron peak frequency and flux needed for the calculation. We use the {\it SDSS} Sky Explorer \footnote{https://tools.ssdc.asi.it/} to plot the SED and fit the synchrotron and inverse-Compton peak. 
In addition, we also calculate the inverse-Compton peak for power-law-like spectra \citep{lea_blazar_compton} using:
\begin{equation} \label{eq2}
\frac{d N}{d E}=K\left[\left(\frac{E}{E_{b}}\right)^{\delta_{1}}+\left(\frac{E}{E_{b}}\right)^{\delta_{2}}\right]^{-1}.
\end{equation}
 The break energy $E_b$ is found by the $E_b-\Gamma$ relationship in \cite{ajello_ep_index}. The K is normalized to the flux integrated from 1 to 100 GeV in the 4FGL catalog.

\subsection{The Fermi Blazar Divide}

The Fermi blazar divide found by \cite{ghisellini2009fermi} separates the blazar population into BL Lac and FSRQ by gamma-ray luminosity ($L_{\gamma}$) and gamma-ray index ($\Gamma_{\gamma}$): BL Lacs dominate the area with the lower gamma-ray luminosity ($L_{\gamma}<10^{47}\ \mathrm{erg\ s^{-1}}$) and harder spectra index ($\Gamma_{\gamma}<$2.2) in the $\Gamma_{\gamma}-L_{\gamma}$ space. Figure \ref{fig:blazar_divide} shows the Fermi blazar divide and the high-$z$ BL Lacs discovered by the photo-$z$ campaign, using the 4LAC catalog data \citep{Ajello_2020}. We utilized the same method in \cite{sheng2024revealing} to calculate the gamma-ray luminosity from 0.1 GeV to 100 GeV. The plot indicates the distinct separation between the FSRQ and BL Lac populations. However, the high-$z$ BL Lacs found by our photometric method stand out from the two groups in the plot: they are luminous BL Lacs with harder spectra. These outliers can be explained by the ``blue quasars'' or ``masquerading BL Lacs'' \citep{ghisellini2012blue, padovani2012discovery, giommi2012simplified, giommi2013simplified}. Masquerading BL Lacs are FSRQs classified as BL Lacs because the luminous synchrotron emission from the jet saturates the emission lines from the host galaxy. 

We also calculated the disk-to-Eddington luminosity ratio following the same assumption and method in \cite{sheng2024revealing}. The four high-$z$ BL Lacs, 4FGL J0506.9$+$0323, 4FGL J1125.1$-$2101, 4FGL J1823.5$+$6858, 4FGL J2144.2$+$3132, have ratios of 0.008, 0.037, 0.039, 0.147 respectively. Three of them show efficient accretion from the disk (ratio $> 0.02$), indicating the fingerprints of FSRQ. Figure \ref{fig:blazar_divide} shows that the photo-$z$ campaign can efficiently reveal potential masquerading BL Lacs. Using the high-$z$ BL Lacs found from \cite{kaur2017} (4FGL J2146.5$-$1344 and 4FGL J1520.8$-$0348), \cite{rajagopal2020nustar} and \cite{rajguru2024xmm} found evidence that the properties of these sources could be explained by the ``blue FSRQ'' scenario.

%Notice that we apply some underlying assumptions when inferring the ratio of $L_{\mathrm{disk}}$ to $L_{\mathrm{Edd}}$. We would suggest further analysis on these sources.

\section{Summary and Conclusions} \label{sec:summary}
This work studied 64 blazars, among which 59 are BL Lacs and 5 are BCU. The sources are observed in ten filters by the {\it Swift}/UVOT and SARA-CT/RM telescopes. The photometric data are produced via the automatic pipeline set up by the {\it photozpy} package. Following the method implemented in \cite{Rau2012}, we obtained redshifts ($z>1.3$) for 4 BL Lacs and upper limits for 50 sources. So far, the photo-$z$ campaign has discovered 23 high-$z$ BL Lacs, including this work. The rare high-$z$ BL Lacs are essential probes for the EBL models and blazar population. %Our sample examines the cosmic gamma-ray horizon: the two high-$z$ sources that lie above the horizon might indicate that the current EBL model might be too opaque. 
In addition, our work also shows that the blazar sequence might be subject to selection effects due to the challenges in the BL Lac redshift measurement, and the missing luminous high-$z$ BL Lacs are found by our method. Moreover, our high-$z$ BL Lacs are potential masquerading BL Lacs due to their high luminosity and hard spectra. %Three high-$z$ sources found in this paper show efficient disk accretion, indicating that they might be FSRQs whose broad-line emission is outshone by the luminous non-thermal emission from the jet.

\acknowledgements

Y.S. and M.A. acknowledge funding under NASA contracts 80NSSC24K0629, 80NSSC23K1150 and 80NSSC22K1472. This work is also supported in part by NSF AST 2319415. The authors acknowledge the \emph{Swift} team for scheduling all the \emph{Swift}/UVOT observations. They also acknowledge the Southeastern Association for Research in Astronomy for providing SARA-CT and SARA-RM observations. A.D. is thankful for the support of the Ramón y Cajal program from the Spanish MINECO.

\software{
   Astropy (\citealp{astropy}),\ 
   HEASoft (v6.28; \citealp{heasoft}),\ 
   numpy (\citealp{numpy}),\ 
   SAO Image DS9 (\citealp{ds9}),\ 
   Matplotlib (\citealp{matplotlib}),\ 
   Pandas (\citealp{pandas}),\ 
   Scipy (\citealp{sciPy}),\
   swifttools\footnote{\url{https://gitlab.com/DrPhilEvans/swifttools}},\
   ccdproc (\citealp{ccdproc}),\
   photutils \citealp{photutils}),\ 
   astroalign (\citealp{BEROIZ2020100384}),\ 
   astroquery (\citealp{astroquery}),\
   regions \footnote{\url{https://github.com/astropy/regions}},\ 
   LePhare \citep{arnouts1999measuring,ilbert2006accurate}.
   }

\bibliographystyle{apj}
\bibliography{SARA_Photo_z}

\end{document}